\documentclass[3p,authoryear,review,11pt,]{elsarticle}

\usepackage{lineno,hyperref}
\usepackage{natbib}
\usepackage{adjustbox}
\usepackage{longtable}
\usepackage{booktabs}
\usepackage{xcolor}
\usepackage{caption}
\usepackage{booktabs}
\usepackage{soul}
\usepackage{textcomp}
\usepackage{multirow}
\usepackage{multicol}
\usepackage{tabu}
\usepackage{tocloft}
\usepackage{blindtext}
\usepackage{mathtools}
\usepackage{blkarray}
\usepackage{subcaption}
\usepackage{float}
\usepackage{caption}
\usepackage{rotating}
\usepackage{lscape}
\usepackage{array}
\usepackage{booktabs}
\usepackage{soul}
\usepackage{textcomp}
\usepackage{amsmath}
\usepackage{amssymb}
\usepackage{setspace}
\usepackage{bm}
\usepackage{makecell}
\usepackage{amsfonts}
\usepackage{algorithm}
\usepackage{algorithmic}
\usepackage{booktabs}
\usepackage{subcaption}
\usepackage{tikz} \usetikzlibrary{positioning, fit, calc, bayesnet}
\tikzset{
  plate/.style={draw, shape=rectangle, rounded corners=0.5ex, thick,
    minimum width=3.1cm, text width=3.1cm, align=right, inner sep=10pt, inner ysep=10pt,label={[xshift=-14pt,yshift=14pt]south east:#1}}
}

\usepackage[thmmarks,amsmath]{ntheorem}
\newtheorem{remark}{Remark}
\newtheorem{definition}{Definition}
\usepackage{bbm}

\biboptions{authoryear}

\begin{document}

\begin{frontmatter}

\title{Spherical Double K-Means: a co-clustering approach for text data analysis}

\author[1]{\corref{cor2}Ilaria Bombelli}
\ead{ilaria.bombelli@istat.it}

\author[2]{Domenica Fioredistella Iezzi}
\ead{stella.iezzi@uniroma2.it}

\author[2]{\corref{cor1} Emiliano Seri}
\ead{emiliano.seri@uniroma2.it}
\cortext[cor1]{Corresponding author}

\author[3]{Maurizio Vichi}
\ead{maurizio.vichi@uniroma1.it}

\address[1]{Italian National Institute of Statistics,
Directorate for Methodology and Statistical Process Design, Via Cesare Balbo 16, 00184, Rome, Italy}

\address[2]{Department of Enterprise Engineering, University of Rome Tor Vergata, Via del Politecnico 1, 00133, Rome, Italy}

\address[3]{Department of Statistics, Sapienza University of Rome, Piazzale Aldo Moro 5, 000185, Rome, Italy}

\begin{abstract}

In text analysis, Spherical K-means (SKM) is a specialized k-means clustering algorithm widely utilized for grouping documents represented in high-dimensional, sparse term-document matrices, often normalized using techniques like TF-IDF. Researchers frequently seek to cluster not only documents but also the terms associated with them into coherent groups. To address this dual clustering requirement, we introduce Spherical Double K-Means (SDKM), a novel methodology that simultaneously clusters documents and terms. This approach offers several advantages: first, by integrating the clustering of documents and terms, SDKM provides deeper insights into the relationships between content and vocabulary, enabling more effective topic identification and keyword extraction. Additionally, the two-level clustering assists in understanding both overarching themes and specific terminologies within document clusters, enhancing interpretability. SDKM effectively handles the high dimensionality and sparsity inherent in text data by utilizing cosine similarity, leading to improved computational efficiency. Moreover, the method captures dynamic changes in thematic content over time, making it well-suited for applications in rapidly evolving fields. Ultimately, SDKM presents a comprehensive framework for advancing text mining efforts, facilitating the uncovering of nuanced patterns and structures that are critical for robust data analysis.

We apply SDKM to the corpus of US presidential inaugural addresses, spanning from George Washington in 1789 to Joe Biden in 2021. 
Our analysis reveals distinct clusters of words and documents that correspond to significant historical themes and periods, showcasing the method's ability to facilitate a deeper understanding of the data. Our findings demonstrate the efficacy of SDKM in uncovering underlying patterns in textual data.

\end{abstract}

\begin{keyword}
Text data \sep Co-clustering \sep Topic modeling \sep Spherical double k-means
\end{keyword}

\end{frontmatter}


\section{Introduction}
\label{sec:intro}

In recent years, the rapid increase in textual data from various sources, such as social networks,
electronic health records, insurance claims, and news outlets, has made effective text analysis
methods more critical than ever. Text clustering, the process of grouping similar texts, is crucial in
this analysis.
Text clustering, which involves grouping similar documents without predefined categories or labels \citep{iezzi12}, plays a pivotal role in uncovering hidden patterns, topics, or clusters within large corpora. This facilitates the exploration and organization of vast text datasets based on inherent similarities, aiding in information retrieval, topic modeling, and data summarization.

Among the most used clustering algorithms, k-means and its variants are valued for their simplicity, scalability, and ability to classify large and unstructured data quickly.
However, traditional k-means clustering has limitations when applied to text data. The algorithm’s reliance on Euclidean distance often fails to accurately represent relationships in high-dimensional spaces, such as those created by text representations like TF-IDF or bag-of-words, ignoring meaningful semantic relationships like synonyms and context. Additionally, the high dimensionality of text data can render Euclidean distance less meaningful, leading to ineffective clustering results. Another challenge is k-means’ sensitivity to initialisation, as its performance heavily depends on the number of clusters and the initial placement of centroids, which can result in unstable or suboptimal outcomes. The algorithm also assumes that clusters are spherical and evenly distributed, a condition rarely met in text data, which often forms non-convex or overlapping clusters. Furthermore, k-means treats text as numerical data, lacking semantic understanding, which can lead to similar words being assigned to different clusters. Lastly, k-means is sensitive to outliers and noise, such as rare words or misspellings, distorting cluster centroids and reducing overall clustering quality. Moreover, text data often exhibit high dimensionality, sparsity, and noise, making the Euclidean distance metric used in standard k-means less suitable. This is because Euclidean distance can overemphasize longer documents with higher term weights, leading to biased clustering results \citep{hornik2012spherical}.

To address these challenges, alternative distance measures like cosine similarity have been employed in text clustering. Cosine similarity assesses the similarity between two vectors based on the cosine of the angle between them \citep{maher2016effectiveness}. 

In this paper, we introduce the Spherical Double K-Means (SDKM) clustering method, an original approach for the simultaneous partitioning of terms and documents. SDKM integrates the principles of double k-means clustering (DKM) \citep{DKMvichi2001} and spherical k-means (SKM) \citep{dhillon2001concept, hornik2012spherical}, tailored specifically for textual data analysis. By incorporating cosine similarity into the co-clustering framework, SDKM effectively handles high dimensionality, sparsity, and noise, facilitating the discovery of meaningful patterns in text corpora.

Co-clustering, or the simultaneous clustering of rows and columns in a term-document matrix, allows for the classification of terms based on the entire set of documents and vice versa \citep{celardo2016multi}. This duality between row and column clusters enhances robustness against the challenges of large, sparse, and non-negative matrices common in text analysis. The interrelated nature of row and column clusters in co-clustering methods fosters a more comprehensive understanding of the underlying data structure.

The proposed SDKM methodology offers a versatile approach with broad applicability in various tasks, including topic modeling, document clustering, sentiment analysis, information retrieval, and document summarization. Its ability to capture the inherent structure of textual data makes it particularly well-suited for extracting meaningful insights from large text corpora.

We demonstrate the effectiveness of SDKM through its application to the US presidential inaugural addresses, a corpus spanning from George Washington in 1789 to Joe Biden in 2021, available in the R package \textsf{quanteda} \citep{quant18}. By analyzing this dataset, we uncover thematic evolutions and patterns in presidential rhetoric over more than two centuries.

The remainder of the paper is structured as follows: Section~\ref{sec:background} provides background on text data processing and reviews relevant clustering methodologies. Section~\ref{sec: method} introduces the SDKM methodology in detail. Section~\ref{sec:simulation} presents the data generation procedure, the pseudo-$F$ index for the selection of the optimal number of clusters, and a simulation study to evaluate the performance of the proposed method. Section~\ref{sec:application} showcases the application of SDKM to the US presidential inaugural addresses. 
In Section~\ref{sec:comparison}, we compare the results achieved with SDKM to those achieved with Double K-Means (DKM) in the proposed application, highlighting their similarities and differences.
Finally, Section~\ref{sec:conclusion} concludes the paper and discusses potential directions for future research.

\section{Background}
\label{sec:background}
\subsection{Addressing text data}

In recent years, the exponential growth of text-based information, ranging from social networks and electronic health records to insurance claims and news outlets, has highlighted the necessity of robust text analysis tools. Unlike structured numerical data, textual data is inherently unstructured, high-dimensional, and prone to significant noise and sparsity \citep{Iezzi2020, Churchill2022}. This complexity demands specialized methods of data representation, preprocessing, and clustering to effectively manage large text corpora \citep{celardo2016multi}.
A foundational step in text analysis is transforming unstructured documents into a numeric format. A widely adopted solution is the \textit{Vector Space Model} (VSM), which represents each document as a vector of term frequencies. Although the VSM is conceptually straightforward and often effective, it introduces complications related to document length and high dimensionality \citep{Aggarwal2012MiningTD}. One way to mitigate these effects is by normalizing term vectors or employing similarity measures such as cosine similarity, which can reduce the bias toward longer documents \citep{hornik2012spherical}.

Before converting documents to numeric vectors, the data undergoes a \textit{preprocessing} phase \citep{cozzolino2022document} that typically includes:
\begin{itemize}
    \item Tokenization: Splitting text into smaller units (tokens), for example at the word level (unigrams, bigrams, etc.).
    \item Stopwords Removal: Eliminating high-frequency but minimally informative words (e.g., ``the'', ``and'', ``or'') to decrease dimensionality.
    \item Pruning: Filtering out terms that occur either extremely often or extremely rarely in the corpus, thus reducing noise.
    \item Stemming or Lemmatization: Stemming reduces words to their root or base form by removing suffixes or prefixes, often using heuristic rules.
Lemmatization reduces words to their canonical form (lemma) based on dictionary definitions and linguistic analysis. For example, ``studies'', ``studying'', ``studied'' are reduced to ``study''.
\end{itemize}
By trimming the vocabulary size and normalizing word variants, these steps significantly reduce dimensionality and noise, thereby improving clustering quality and computational efficiency.

Nonetheless, text data remains difficult to analyze due to its high sparsity: most cells in a term-document matrix are zero, and the vocabulary can easily scale to tens or hundreds of thousands of terms. As a result, standard clustering methods may become infeasible or yield poor performance. Further complicating matters, the same concept may appear in numerous synonyms or domain-specific slang, making purely frequency-based techniques inadequate for capturing deeper semantic relationships \citep{lodhi2002text}. Document length is also highly variable; some texts may contain only a few words, whereas others include thousands. Traditional clustering approaches relying on Euclidean metrics thus risk overweighting longer documents \citep{dhillon2001concept}, so cosine-based measures often prove more appropriate for text-based scenarios \citep{hornik2012spherical, zhao2004empirical}.

Many mainstream clustering algorithms have been adapted or extended to reduce the impact of high dimensionality and sparsity in text:
\begin{itemize}
    \item Prototype-based methods (e.g., Spherical K-means): Modify k-means to incorporate cosine similarity, improving results on high-dimensional text corpora \citep{dhillon2001concept}.
    \item Graph-based methods (e.g., spectral clustering): Represent documents as nodes in a graph, with edges weighted by kernel or substring-based similarities \citep{lodhi2002text}.
    \item Hierarchical approaches (e.g., agglomerative, divisive): Build nested clusters \citep{steinbach2000comparison}, which is especially beneficial for smaller corpora or when interpretable dendrograms are desired, though scalability can be an issue.
    \item Model-based approaches (e.g., Latent Dirichlet Allocation): Assume documents arise from mixtures of underlying distributions. While Gaussian mixture models must address text’s sparsity and manifold structure \citep{Bouveyron2019Jun}, \textit{Latent Dirichlet Allocation (LDA)} specializes in revealing latent topics by modeling each document as a mixture of hidden themes \citep{blei2003latent}.
\end{itemize}

\subsection{Spherical k-means and double k-means methodologies}
\label{literature review}

Our methodological proposal integrates two well-known clustering techniques: the former, DKM, is an extension of KM and it is used to simultaneously cluster units and variables of a data matrix; the latter, SKM, is a useful tool to cluster units of a data matrix, as the well-known KM algorithm does; however, the main difference between SKM and KM lies in the fact that SKM uses a non-euclidean distance to compute dissimilarities between any pair of units. In the following a formal definition and detailed description on the two tecnhiques are provided. 

\subsubsection{Double k-means: DKM}

Given a data matrix $\bm{X}=\{x_{ij}:i=1,\dots,N, j=1,\dots,J\}$ where $i$ indicates units and $j$ indicates variables, the DKM algorithm simultaneously partitions units into $K$ clusters and variables into $Q$ clusters. Therefore, the final outputs of the algorithm are two membership matrix modeling the partitions of units and variables into clusters, and a centroids matrix of dimension $K\times Q$, synthesizing the subsets of $\bm{X}$. More in details, the whole data matrix is partitioned into subsets (blocks) and each block is represented by one centroid. The DKM model is formally defined as follows:

\begin{equation}
\label{DKM_model}
\bm{X} = \bm{U}\bm{\overline{Y}}\bm{V}^\prime + \bm{E} 
\end{equation}
Subject to the constraints:
\begin{align*} 
u_{ik} &\in \{0,1\}, \quad \forall i = 1, \dots, N;, k = 1, \dots, K \\ 
\sum_{k=1}^K u_{ik} &= 1, \quad \forall i = 1, \dots, N \\
v_{jq} &\in \{0,1\}, \quad \forall j = 1, \dots, J;, q = 1, \dots, Q \\
\sum_{q=1}^Q v_{jq} &= 1, \quad \forall j = 1, \dots, J  
\end{align*}
where $\bm{U}$ is the $N\times K$ membership matrix modelling the partition of the $N$ units inside $K$ clusters; $\bm{V}$ is the $J\times Q$ membership matrix modelling the partition of the $J$ variables inside $Q$ clusters; $\bm{\overline{Y}}$ is the $K\times Q$ centroids matrix, whose element $\overline{y}_{kq}$ synthesizes the observations within the block identified by the units belonging to the $k$-th cluster and the variables belonging to the $q$-th cluster.

The centroid matrix $\bm{\overline{Y}}$, thus synthesizes the most relevant information of the data matrix $\bm{X}$ and can be considered as a reduced data matrix, with rank at most equal to the minimum between $K$ and $Q$. 
The matrix $\bm{E}$ is the $N\times J$ matrix of errors.
Finally, the constraints $u_{ik} \in \{0,1\}$ and $\sum_{k=1}^K u_{ik} = 1$ ensure that $\bm{U}$ is a binary and row-stochastic matrix; similarly, the constraints $v_{jq} \in \{0,1\}$ and $\sum_{q=1}^Q v_{jq} = 1$ ensure that $\bm{V}$ is a binary and row-stochastic matrix. 

\begin{remark}
    The centroids matrix is updated as follows: 
    \begin{equation}
        \bm{\overline{Y}}=\bm{U}^{+}\bm{X}\bm{V}^{+\prime}=(\bm{U}^\prime \bm{U})^{-1}\bm{UXV}(\bm{V}^\prime \bm{V)}^{-1}
    \end{equation}
    where $\bm{U}^{+}=(\bm{U}^\prime \bm{U})^{-1}\bm{U}$ and $\bm{V}^{+}=(\bm{V}^\prime \bm{V)}^{-1}\bm{V}^{\prime}$ denotes the Moore-Penrose inverse of the matrices $\bm{U}$ and $\bm{V}$, respectively. 

\end{remark}
\begin{remark}
If the units' membership matrix $\bm{U}$ degenerates into an identity matrix of order $K$, i.e. $\bm{U}=\bm{1}_K$, the model \ref{DKM_model} can be written as $\bm{X}=\bm{\overline{Y}V}^{\prime}+\bm{E}$, which is the model of the KM to partition variables into $Q$ clusters. 
\newline Similarly, if the variables' membership matrix $\bm{V}$ degenerates into an identity matrix of order $J$, i.e. $\bm{V}=\bm{1}_J$, the model \ref{DKM_model} can be written as $\bm{X}=\bm{U\overline{Y}}+\bm{E}$, which is the model of the KM to partition units into $K$ clusters. 

\end{remark}

\subsubsection{Spherical k-means: SKM}

Given $\bm{X}$, SKM partitions units into $K$ clusters. First proposed by \cite{dhillon2001concept} and then specified in \cite{hornik2012spherical}, what makes it different from the KM algorithm, whose objective is the same, is the use of the distance function to compute dissimilarities between any pair of units: indeed, the standard KM use Euclidean distance, while the SKM computes \textit{cosine dissimilarity}. 
\begin{definition}
    Given two vectors $\bm{a}$ and $\bm{b}$, the \textit{cosine dissimilarity} between them is defined as follows: 
    \begin{equation}\label{cosineD}
        \text{d}(\bm{a}, \bm{b})=1-cos(\bm{a},\bm{b})=1-\frac{\langle \bm{a},\bm{b} \rangle}{\lVert \bm{a} \rVert \lvert \bm{b} \rVert}
    \end{equation}
    The cosine dissimilarity takes into account the angle between the two vectors. 
    \newline If the two vectors $\bm{a}$ and $\bm{b}$ are normalized, i.e. $ \lvert a \rVert =  \lvert b \rVert= 1$, then the cosine dissimilarity becomes: 
    \begin{equation}\label{cosineDnorm}
        \text{d}(\bm{a}, \bm{b})=1-cos(\bm{a},\bm{b})=1-\langle \bm{a},\bm{b} \rangle
    \end{equation}
    Cosine dissimilarity takes value 0 if the vectors are exactly equal, while it takes value 1 if the two vectors do not share any elements.
\end{definition}
The SKM algorithm aims to minimize the (cosine) distance between observations and centroids. Therefore, the objective function of the SKM is formalized as follows: 

\begin{equation}\label{objfunction_SKM}
f(\bm{U},\bm{\overline{X}})=\sum_{i=1}^N \sum_{k=1}^K u_{ik}(1-cos(\bm{x}_i, \bm{\overline{x}_k}))=\sum_{i=1}^N \sum_{k=1}^K u_{ik}\Big(1-\frac{\langle \bm{x_i},\bm{\overline{x_k}} \rangle}{\lVert \bm{x_i} \rVert \lvert \bm{\overline{x}_k} \rVert}\Big)
\end{equation}
s.t.
\begin{align*}
    u_{ik} &\in\{0,1\}, \quad \forall\ i=1,\dots,N; k=1,\dots, K \\
    \sum_{k=1}^K u_{ik}&=1, \quad \forall\ i=1,\dots,N
\end{align*}
where $u_{ik}\in\{0,1\}$ is the membership of unit $i$ to cluster $k$ and $\bm{\overline{x}_k}$ is the $k$-th centroid, synthesising the observations (units) belonging to the $k$-th cluster. 
\begin{remark}
    Likewise, instead of minimizing \ref{objfunction_SKM}, the following objective function can be maximized:
\begin{equation}\label{objfunction_SKM_max}
f(\bm{U},\bm{\overline{X}})=\sum_{i=1}^N \sum_{k=1}^K u_{ik}s_{ik}=\sum_{i=1}^N \sum_{k=1}^K u_{ik}(cos(\bm{x}_i, \bm{\overline{x}_k}))=\sum_{i=1}^N \sum_{k=1}^K u_{ik}\Big(\frac{\langle \bm{x_i},\bm{\overline{x_k}} \rangle}{\lVert \bm{x_i} \rVert \lvert \bm{\overline{x}_k} \rVert}\Big)
\end{equation}
The function in Equation \ref{objfunction_SKM_max} is based on the \textit{cosine similarity} of the angle between observation $\bm{x}_i$ and centroid $\bm{\overline{x}}_k$, i.e. $s_{ik}=cos(\bm{x}_i, \bm{\overline{x}_k})=\Big(\frac{\langle \bm{x_i},\bm{\overline{x_k}} \rangle}{\lVert \bm{x_i} \rVert \lvert \bm{\overline{x}_k} \rVert}\Big)$
\end{remark}
\par The objective function in Equation \ref{objfunction_SKM_max} can be also written in a matrix form notation.

Let $\bm{M}$ be a $N\times N$ diagonal matrix having on the main diagonal the inverse of the norm of the rows of the data matrix $\bm{X}$, defined as follows: 
\begin{equation*}
\label{inverseNorm1}
\bm{M}=\begin{pmatrix}
    \frac{1}{\lVert{\bm{x}_1}\rVert} & 0 & \hdots & 0 \\
    0 &  \frac{1}{\lVert{\bm{x}_2}\rVert} & \ddots & \vdots \\
    \vdots & \ddots & \ddots & 0 \\
    0 & \hdots & \hdots &  \frac{1}{\lVert{\bm{x}_N}\rVert}
\end{pmatrix}
=(diag(\bm{XX^\prime}))^{-\frac{1}{2}}
\end{equation*}

Let $\bm{W}$ be a $K\times K$ diagonal matrix having on the main diagonal the inverse of the norm of the rows of the centroid data $\bm{\overline{X}}$: 
\begin{equation*}\label{inverseNorm2}
\bm{W}=\begin{pmatrix}
   \frac{1}{\lVert{\bm{\overline{x}}_1}\rVert} & 0 & \hdots & 0 \\
    0 &  \frac{1}{\lVert{\bm{\overline{x}}_2}\rVert} & \ddots & \vdots \\
    \vdots & \ddots & \ddots & 0 \\
    0 & \hdots & \hdots &  \frac{1}{\lVert{\bm{\overline{x}}_K}\rVert}
\end{pmatrix}
=(diag(\bm{\overline{X}\overline{X}^}\prime))^{-\frac{1}{2}}
\end{equation*}
Then, the objective function \ref{objfunction_SKM_max} to be maximized can be reformulated in matrix form as follows: 
\begin{equation*}\label{objfunction_SKM_max_matrix}
f(\bm{U},\bm{\overline{X}})=tr(\bm{MX\overline{X}}^\prime \bm{W}\bm{U}^\prime )
\end{equation*}
where $tr(\cdot)$ is the trace operator. It has to be noted that $\bm{S}=\bm{X\overline{X}}^\prime \bm{U}^\prime$ is the $N\times N$ matrix modelling similarities between the data matrix $\bm{X}$ and the centroids matrix $\bm{\overline{X}}$ and having on the main diagonal the similarities between the $N$ observations and their related centroids. More in detail, given a unit $i$ belonging to cluster $k$, then $s_{ik}$ denotes the similarity between observation $\bm{x}_i$ and the centroid $\bm{\overline{x}}_k$.

If the data matrix $\bm{X}$ and the centroids matrix $\bm{\overline{X}}$ are normalized by row, then the objective function becomes: 
\begin{equation*}\label{objfunction_SKM_max_matrix_normalized}
f(\bm{U},\bm{\overline{X}})=tr(\bm{X\overline{X}}^\prime \bm{U}^\prime ) = tr(\bm{X}^\prime \bm{U}\bm{\overline{X}})
\end{equation*}
where the last equality is given by the trace operator property $tr(\bm{AB}^\prime)=tr(\bm{A}^\prime \bm{B})$. 
The normalized version of the objective function, i.e. an objective function which is upper bounded by 1, is the following:
\begin{equation*}
f(\bm{U},\bm{\overline{X}})= \frac{tr(\bm{X}^\prime \bm{U}\bm{\overline{X}})}{\sqrt{tr(\bm{X}^\prime \bm{X})\cdot tr(\bm{\overline{X}}^\prime \bm{U}^\prime \bm{U}\bm{\overline{X}})}}
\end{equation*}
For the sake of notation, we let $\bm{X}_t=\bm{U\overline{X}}$ be the theoretical matrix of the SKM model. Then, the objective function can be written in a more compact way as follows:
\begin{equation}\label{objfunction_SKM_max_matrix_normalized_final}
f(\bm{X}_t)=\frac{tr(\bm{X^\prime X}_t)}{\sqrt{tr(\bm{X^\prime X})\cdot tr(\bm{X}_t^\prime  \bm{X}_t})}
\end{equation}
It has to be noted that the objective function in Equation \ref{objfunction_SKM_max_matrix_normalized_final} is upper bounded by 1 and it is equal to 1 when the data matrix $\bm{X}$ matches the theoretical one $\bm{X}_t$.

In the following, the updating formula of the centroids matrix $\bm{\overline{X}}$ and partition $\bm{U}$ will be derived (Equations \ref{updating_centroids_SKM} and \ref{updating_partition_SKM}) and we will assume that the data matrix $\bm{X}$ and the centroids matrix $\bm{\overline{X}}$ are normalized by row. 
In order to maximize the objective function, it can be useful to consider the problem row-wise: for every row, it is needed to maximize the similarity between observation $\bm{x}_i$ and observation $\bm{\overline{x}}_k$, where $k$ denotes the cluster to which unit $i$ belongs. As it is usually done when maximizing a function, the maximizer can be found by setting equal to zero the first derivative, find the root of the equation, and  then verify that the root is indeed a maximizer. Let $\theta$ denote the angle between $\bm{x}_i$ and $\bm{\overline{x}}_k$, let $f(\theta)=cos(\theta)$ be the function to be maximized. Then, $\frac{df}{d\theta}=-sin(\theta)=0 \iff sin(\theta)=0$.
Therefore, we can conclude that the objective function is maximized when the angle $\theta$ between $\bm{x}_i$ and $\bm{\overline{x}}_k$ is a zero angle or a straight angle, namely when the two vectors  $\bm{x}_i$ and $\bm{\overline{x}}_k$ coincide or are parallel to each other. Formally, the function is maximized when $\bm{x}_i=c\cdot \bm{\overline{x}}_k, c\in\mathbb{R}$.

By extending the reasoning to the whole data matrix $\bm{X}$, 
the objective function is maximized when each of the row-vectors of the data matrix $\bm{X}$ is proportional to the centroids row-vector $\bm{\overline{x}}_k$, where $k$ is the cluster unit $i$ belongs to. Formally, the objective function is maximized when 
\begin{equation}\label{updatingSKM_1}
    \bm{X}=\bm{CU\overline{X}}
\end{equation}
where $\bm{C}$ is an $N\times N$ diagonal matrix whose diagonals elements are the constants of proportionality between each pair of vectors.
\par By solving the Equation \ref{updatingSKM_1} w.r.t. $\bm{\overline{X}}$, we obtain 
\begin{equation*}\label{updatingSKM_2} \bm{U}^\prime\bm{X}=\bm{U}^\prime\bm{CU\overline{X}} \iff \bm{\overline{X}}=\bm{(U}^\prime \bm{CU)^{-1}U}^\prime \bm{X}
\end{equation*}
where firstly we pre-multiply both sides of the equation by $\bm{U}^\prime$ to make $\bm{U}^\prime\bm{C}\bm{U}$ invertible and then we isolate $\bm{\overline{X}}$.
\par In addition, since the centroids matrix $\bm{\overline{X}}$ must be normalized by row, we obtain the following updating formula for $\bm{\overline{X}}$: 
\begin{equation}\label{updating_centroids_SKM}
    \bm{\overline{X}}=\frac{\bm{(U}^\prime \bm{U)}^{-1}\bm{U}^\prime \bm{X}}{\lVert \bm{(U}^\prime \bm{U)^{-1}U}^\prime \bm{X} \rVert}
\end{equation}

Finally, the partition of units inside clusters is provided by including the generic unit $i$ into the cluster whose centroid is the closest one according to the cosine similarity.  Formally, the partition of units inside cluster $k$ is 
\begin{equation}\label{updating_partition_SKM}
\pi_k=\{i\ \text{s.t.}\ \bm{x}_i^\prime \bm{\overline{x}}_k \leq \bm{x}_i^\prime \bm{\overline{x}}_l, \ \forall\ l \in \{1,\dots,K   \},l\neq k\}   \hspace{1cm} \forall\ k\in\{1, \dots, K\}
\end{equation}


\subsubsection{Spherical K-Means for clustering textual data}
The main difference between KM and SKM lies in the choice of metric used. When evaluating the distance between two vectors using the Euclidean metric, high values can result even when the vectors have the same direction and sense but very different magnitudes. This issue does not occur with cosine similarity, which considers the angle between the vectors rather than their length.
This is a crucial difference when conducting cluster analysis on textual documents, as different documents can vary in length. Consequently, the frequency of a term within them also depends on their length. With cosine similarity, two documents with terms appearing in proportionate amounts are considered equivalent, unlike with the Euclidean metric.
Using cosine similarity, two documents with no terms in common will have a similarity of zero due to the orthogonality of the vectors. Documents that share a similar vocabulary will have a high similarity value.

\section{Methodology: Spherical double k-means}
\label{sec: method}

We integrate SKM in DKM  to simultaneously cluster units and variables with spherical shape clusters. We therefore introduce the Spherical double k-means (SDKM).
SDKM methodology aims to incorporate the advantages of SKM, i.e. its ability to address over-dispersion in the data and detect noise, into a double clustering scenario.

Suppose we want to partition using $K$ clusters for row profiles and $Q$ clusters for column profiles. Let $\bm{U}_{N\times K}$ be the matrix of memberships for the row profiles, $\bm{V}_{J\times Q}$ the matrix of memberships for the column profiles, and $\bm{\overline{Y}}_{K\times Q}$ the matrix of centroids. Assuming the goal is to find an approximation matrix $\bm{X}_t$ of $\bm{X}$, such that:
\begin{equation*}
    \bm{X} = \bm{X}_t + \bm{E}
\end{equation*}

where $\bm{X}_t$ is derived from the decomposition of matrix $\bm{X}$ and is defined as:
\begin{equation*}
    \bm{X}_t = \bm{U}\bm{\overline{Y}}\bm{V}'
\end{equation*}

The cosine similarity between matrices $\bm{X}$ and $\bm{X}_t$ can be formulated as:
\begin{equation*}
    \text{tr}(\bm{X}'\bm{X}_t) = \text{tr}(\bm{X}'\bm{U\overline{Y}V}') = \text{tr}(\bm{XV \overline{Y}U}') = \text{tr}(\bm{XX}_{t}')
\end{equation*}
The objective is then to maximize this similarity function by solving the following maximization problem:
\begin{equation*}
    \max_{\bm{U}, \bm{V}, \bm{\overline{Y}}} f(\bm{U}, \bm{V}, \bm{\overline{Y}}) = \text{tr}(\bm{X}'\bm{U}\bm{\overline{Y}}\bm{V}')
\end{equation*}

As in the case of SKM, the centroid matrix that maximizes the objective function is derived by maximizing the cosine of the angle formed by each row of matrix $\bm{X}$ and the approximated data matrix $\bm{X}_t$,  which is maximized when the two vectors are equal or proportional to each other, i.e., when:
\begin{equation*}
   \bm{X} = \bm{C} \bm{U} \bm{\overline{Y}} \bm{V}'
\end{equation*}

where $\bm{C}$ is a diagonal matrix containing the proportionality constants. Solving this yields the expression for the centroid matrix $\bm{\overline{Y}}$:
\begin{equation*}
    \bm{\overline{Y}} = (\bm{U}'\bm{C}\bm{U})^{-1} \bm{U}'\bm{X}\bm{V} (\bm{V}'\bm{V})^{-1}
\end{equation*}

Since matrix $\bm{\overline{Y}}$ must be normalized by rows, it is given by:
\begin{equation*}
\bm{\overline{Y}} = \frac{(\bm{U}'\bm{U})^{-1} \bm{U}'\bm{X}\bm{V} (\bm{V}'\bm{V})^{-1}}{\| (\bm{U}'\bm{U})^{-1} \bm{U}'\bm{X}\bm{V} (\bm{V}'\bm{V})^{-1} \|}
\end{equation*}

The objective function of the Double Spherical K-means can be normalized:
\begin{equation*}
 \max_{\bm{U}, \bm{V}, \bm{\overline{Y}}} f(\bm{U}, \bm{V}, \bm{\overline{Y}}) = \frac{\text{tr}(\bm{X}'\bm{U}\bm{\overline{Y}}\bm{V}')}{\sqrt{\text{tr}(\bm{X}'\bm{X})\text{tr}(\bm{U\overline{Y}V}'\bm{V\overline{Y}}'\bm{U}'})}
\end{equation*}

In this way, the objective function $\max\limits_{\bm{U}, \bm{V}, \bm{\overline{Y}}} f(\bm{U}, \bm{V}, \bm{\overline{Y}})$ is less than or equal to 1, and it assumes the value 1 when the data matrix $\bm{X}$ coincides with the approximated matrix $\bm{X}_t$. 
It can be observed that if the membership matrix of the variables $\textbf{V}$ degenerates into an identity matrix of order $J$, the SDKM problem reduces to the SKM problem concerning the rows. In this case, the objective function becomes: 
\begin{equation*}
    \max_{\bm{X}, \bm{Y}} \text{tr}(\bm{X}' \bm{\overline{Y}} \bm{V}')
\end{equation*}
and the centroid matrix becomes $(\bm{\overline{Y}}=\bm{U}'\bm{X})$. Similarly, if the membership matrix of the units \( \bm{U} \) degenerates into an identity matrix of order \( N \), it is necessary to cluster only the variables of the data matrix, and the maximization problem of the SDKM is reduced to the following:

\begin{equation*}
\max_{\bm{X}, \bm{Y}} \text{tr}(\bm{X}' \bm{\overline{Y}} \bm{V}')    
\end{equation*}

where the centroid matrix is \( \bm{\overline{Y}} = \bm{X} \bm{V} \).

The Spherical Double K-means algorithm proceeds as follows:

\begin{algorithm}[H]
\caption{Double Spherical K-means Algorithm}
\begin{algorithmic}[1]
\STATE \textbf{Initialization:} Randomly select membership matrices $\bm{U}$ and $\bm{V}$, set iteration index $t = 0$.
\REPEAT
    \STATE \textbf{Step 1:} Update the membership matrix $\bm{U}$.
    \FOR{$i = 1$ to $N$}
        \STATE $u_{ik} = \begin{cases}
            1 & \text{if } x_i (\bm{\overline{Y}} \bm{V}^\prime) I_k = \max\limits_{l} x_i (\bm{\overline{Y}} \bm{V}^\prime) I_l \\
            0 & \text{otherwise}
        \end{cases}$
    \ENDFOR
    \STATE \textbf{Step 2:} Update the centroids.
    \STATE $\bm{\overline{Y}} = \dfrac{(\bm{U}^\prime \bm{U})^{-1} \bm{U}^\prime \bm{X} \bm{V} (\bm{V}^\prime \bm{V})^{-1}}{\left\| (\bm{U}^\prime \bm{U})^{-1} \bm{U}^\prime \bm{X} \bm{V} (\bm{V}^\prime \bm{V})^{-1} \right\|}$
    \STATE $t \leftarrow t + 1$
    \STATE \textbf{Step 3:} Update the membership matrix $\bm{V}$.
    \FOR{$j = 1$ to $J$}
        \STATE $v_{jq} = \begin{cases}
            1 & \text{if } x_j^\prime (\bm{U} \bm{\overline{Y}}) \bm{W}_q = \max\limits_{l} x_j^\prime (\bm{U} \bm{\overline{Y}}) \bm{W}_l \\
            0 & \text{otherwise}
        \end{cases}$
    \ENDFOR
    \STATE \textbf{Step 4:} Update the centroids again.
    \STATE $\bm{\overline{Y}} = \dfrac{(\bm{U}^\prime \bm{U})^{-1} \bm{U}^\prime \bm{X} \bm{V} (\bm{V}^\prime \bm{V})^{-1}}{\left\| (\bm{U}^\prime \bm{U})^{-1} \bm{U}^\prime \bm{X} \bm{V} (\bm{V}^\prime \bm{V})^{-1} \right\|}$
    \STATE $t \leftarrow t + 1$
\UNTIL{Stopping criterion is met}
\end{algorithmic}
\end{algorithm}

As a stopping rule, the objective function $f(\bm{U}, \bm{V}, \bm{\overline{Y}})$ must be computed. If the increase in the objective function is smaller than a predefined tolerance $\epsilon$, the algorithm is considered to have converged; otherwise, return to Step 1.

Here, $\bm{I}$ and $\bm{W}$ denote the identity matrices of dimensions $K$ and $Q$, respectively. The above algorithm guarantees that the objective function's value increases monotonically at each iteration.

Indeed, it can be shown that for any \( t \geq 0 \):

\begin{equation*}
f(\bm{U}^{(t)}, \bm{\overline{Y}}^{(t)}, \bm{V}^{(t)}) \leq f(\bm{U}^{(t+1)}, \bm{\overline{Y}}^{(t+1)}, \bm{V}^{(t+1)})
\end{equation*}

Letting \( p_k^{(t)} \) denote the centroid of the generic cluster \( k \) of the units, i.e., rows, at step \( t \) of the algorithm, with \( 1 \leq k \leq K \), and \( \pi_k^{(t)} \) the corresponding cluster, the previous inequality is demonstrated in Appendix~\ref{appendix:demonstrations}.

\subsection{Normalization issue}

In the context of spherical clustering, it is essential for the centroids to lie on the unit hypersphere to properly measure cosine similarity. This requires the centroid matrix $\bm{\overline{Y}}$ to be normalized. However, in a co-clustering framework where we simultaneously cluster both rows and columns, achieving normalization in both dimensions presents a challenge. Normalizing $\bm{\overline{Y}}$ by both rows and columns is not feasible due to the interdependence between the clusters of units and variables.

To address this issue, we choose to normalize the centroid matrix $\bm{\overline{Y}}$ by rows. This decision is based on the consideration that in textual data, terms (represented by rows in $\bm{X}$) typically exhibit more variability and noise than documents (columns). By normalizing the centroids by rows, we ensure that each centroid vector for the term clusters has a unit norm, aligning with the principles of spherical clustering. This enhances the method's robustness against noise in the term dimension, which is crucial for text data where term frequencies can vary widely.
While our approach does not normalize the columns of $\bm{\overline{Y}}$, we observe that the column norms remain close to one, which suggests that the lack of column normalization may not significantly impact the clustering results. 

\section{Cluster validity and simulation study}
\label{sec:simulation}

The simulation study has been developed in order to test the model and algorithm's performance. 
We implemented a data generation procedure in order to generate datasets whose clustering structure follows a SDKM model. 
The generated datasets are then used to focus on different task: (i) development of the pseudo-$F$ index, proposed by \cite{vichi2015two}, to detect the true number of clusters of units $K$ and the true number of clusters of variables $Q$; (ii) implementation of different scenarios (corresponding to different error levels) in order to assess model performance in terms of true partitions recovery, measured with the Adjusted Rand Index (ARI) \citep{hubert1985comparing}, and centroids matrix recovery, measured with Rooted Mean Squared Error (RMSE) and its Normalized versions (NRMSE1, NRMSE2). Clearly, there exist several normalization techniques: we decided to implement NRMSE1, where the RMSE is normalized by dividing it by the range of the true centroids matrix; NRMSE2, instead, is obtained by dividing RMSE by the norm of centered centroids matrix.

\subsection{Data Generation Mechanism}
We implemented a MATLAB procedure which allows to generate a data matrix having clustering structure following SDKM model. 
\par The data matrix $\bm{X}$ is generated as follows:
\begin{equation*}\label{data_gen}
\bm{X}=\bm{U}\bm{\overline{Y}}\bm{V}^\prime +\bm{E}    
\end{equation*}
where $\bf{U}$ is the (N $\times$ K) membership matrix of units;  $\bf{V}$ is the (J $\times$ Q) membership matrix of variables; $\bm{\overline{Y}}$ is the (K $\times$ Q) matrix of centroids; $\bf{E}$ is the (N $\times$ J) matrix of errors, generated from a normal distribution centered in 0. The membership matrices have been generated by randomly permuting the rows of an initial matrix, where the first 
rows form an identity matrix and the remaining rows contain randomized binary vectors with a 1 in the first column, ensuring variability while maintaining a structured pattern; the matrix of centroids has been generated in such a way centroids are equidistant: a regular (10)-dimensional simplex has been generated; then, a multidimensional scaling has been applied; finally, the centroid matrix is obtained by multiplying the largest real eigenvectors and the square root of the largest real eigenvalues. This procedure allows to have centroids which are equidistant.
\par In the generation mechanism, we introduced two sources of error to simulate heterogeneity in the data:
\begin{itemize}
    \item Centroid Error (\( \epsilon_{\text{centroid}} \)): This error term adds variability to the centroids of the clusters, simulating the effect of noise within clusters.
    \item Cluster Error (\( \epsilon_{\text{cluster}} \)): This error term introduces variability between clusters, affecting the distinctness of the cluster separation.
\end{itemize}

\subsection{Selection of the number of clusters}

To determine the appropriate numbers of clusters K and Q, we employed the pseudo-$F$ index proposed by \cite{ROCCI20081984}. This index, which is based on the criterion proposed by \cite{Calinski1974}, has previously been utilized for the DKM algorithm.

\begin{equation*}
pF_{dk}=\frac{\left\|\bm{H_{U}XH_{V}}-(1/NJ)\bm{1_{N}1_{N}}^{\prime}\right\|^{2}/(KQ-1)}{\left\|\bm{X}-\bm{H_{U}XH_{V}}\right\|^{2}/(NJ-KQ)}.
\end{equation*}
In this equation, $\bm{H_U}$ and $\bm{H_V}$ are projection matrices that map the data onto the subspaces defined by the unit and feature clusters, respectively, and $\bm{1_{N}}$ is a vector of ones of length N. The numerator represents the between-cluster deviance, while the denominator accounts for the within-cluster deviance, with each term normalized by its corresponding degrees of freedom. 

The optimal values of K and Q are identified by locating the local maximum of the $pF_{dk}$ index across different SDKM solutions. As noted by \cite{ROCCI20081984}, this index is effective in scenarios where the data exhibit a well-defined cluster structure.

\subsection{Pseudo-F index simulation study}
\label{subsec: pseudo-F simulation}

We conducted a simulation study to evaluate the effectiveness of the pseudo-$F$ index in selecting the correct number of clusters K and Q under varying levels of error. 
In our simulations, we generated synthetic data with known cluster structures. The data matrix \(\bm{X}\) was constructed based on the true cluster configurations, with \( K_{\text{true}} = 4 \) clusters for units and \( Q_{\text{true}} = 3 \) clusters for variables. 

\begin{figure}[H]
    \centering
    \caption{Heatmaps of the data at different error levels $\epsilon$}
    \label{fig:heatmaps}
    \begin{subfigure}[b]{0.32\textwidth}
        \centering
    \includegraphics[width=\textwidth]{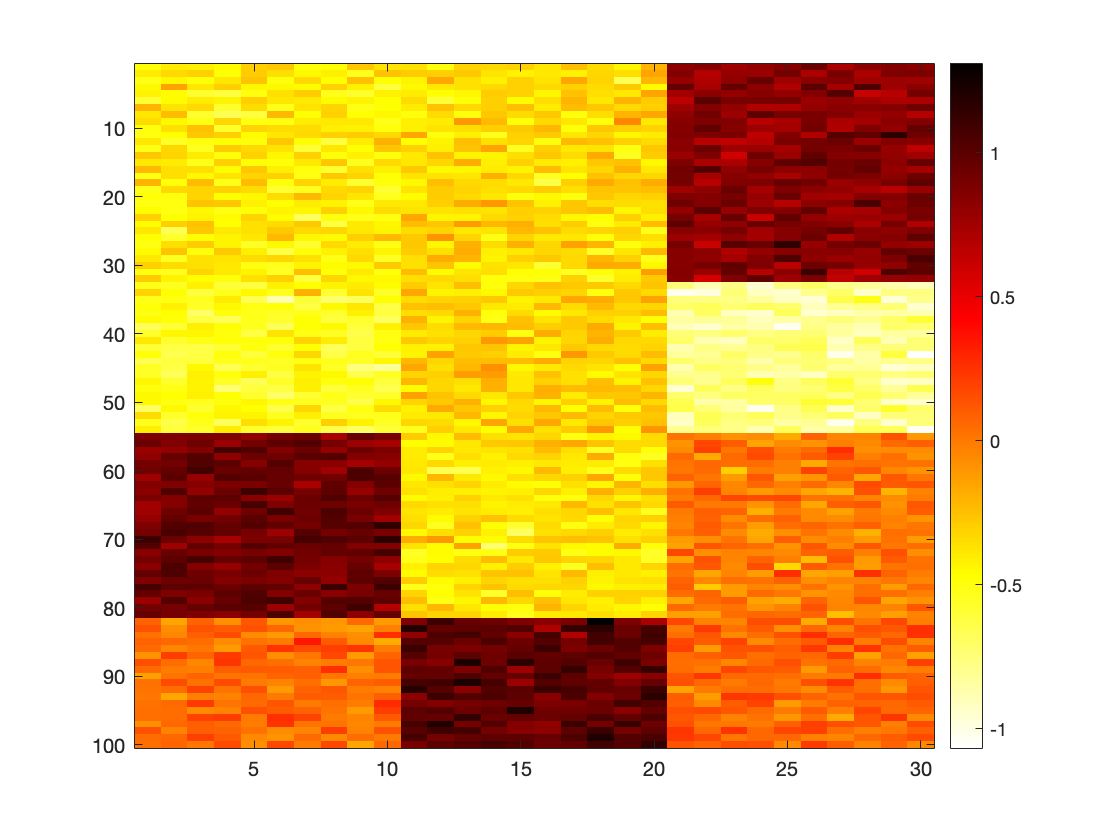}
        \caption{$\epsilon = 0.1$}
    \end{subfigure}
    \hfill
    \begin{subfigure}[b]{0.32\textwidth}
        \centering
    \includegraphics[width=\textwidth]{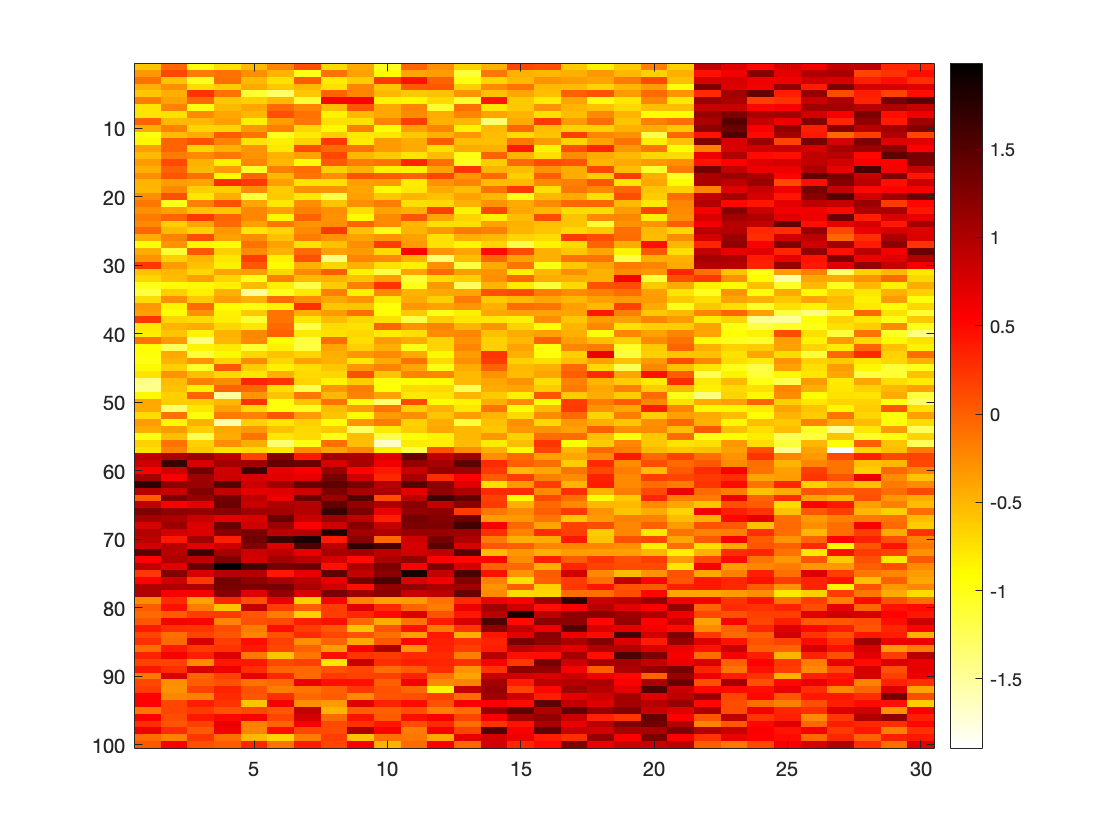}
        \caption{$\epsilon = 0.35$}
    \end{subfigure}
    \hfill
    \begin{subfigure}[b]{0.32\textwidth}
        \centering
    \includegraphics[width=\textwidth]{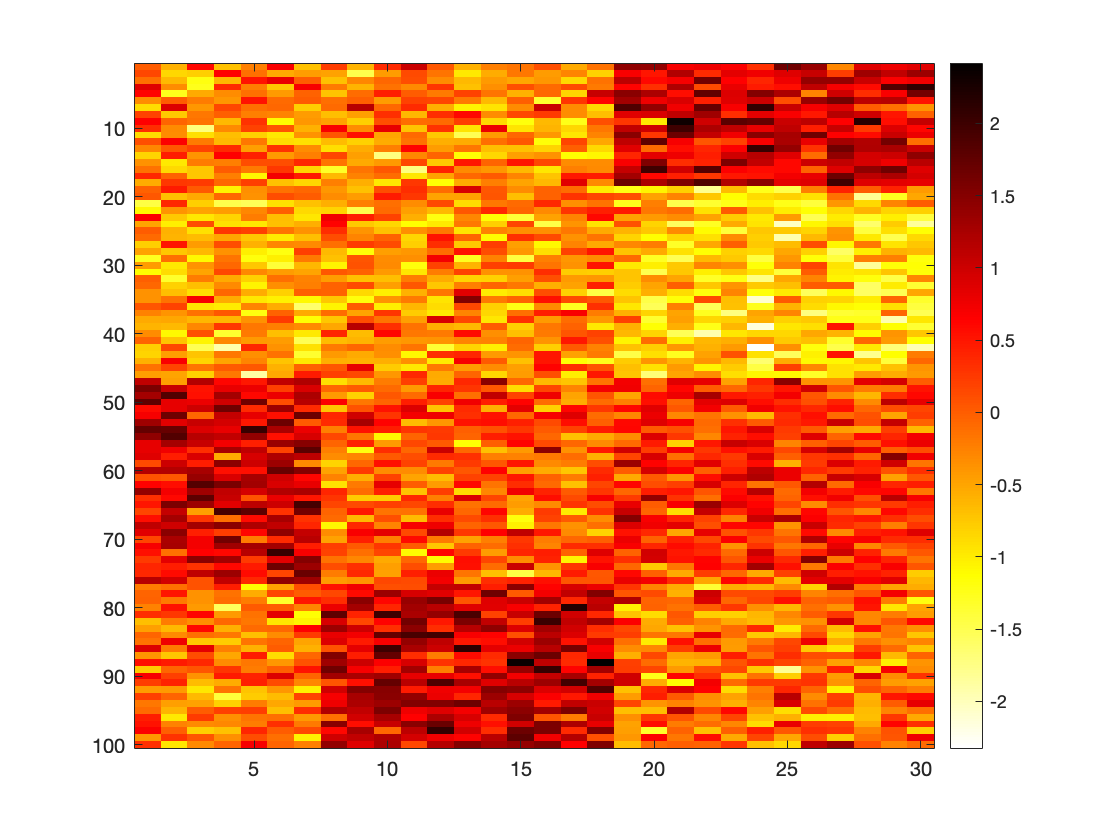}
        \caption{$\epsilon = 0.5$}
    \end{subfigure}
    \vskip\baselineskip 
    \begin{subfigure}[b]{0.32\textwidth}
        \centering
    \includegraphics[width=\textwidth]{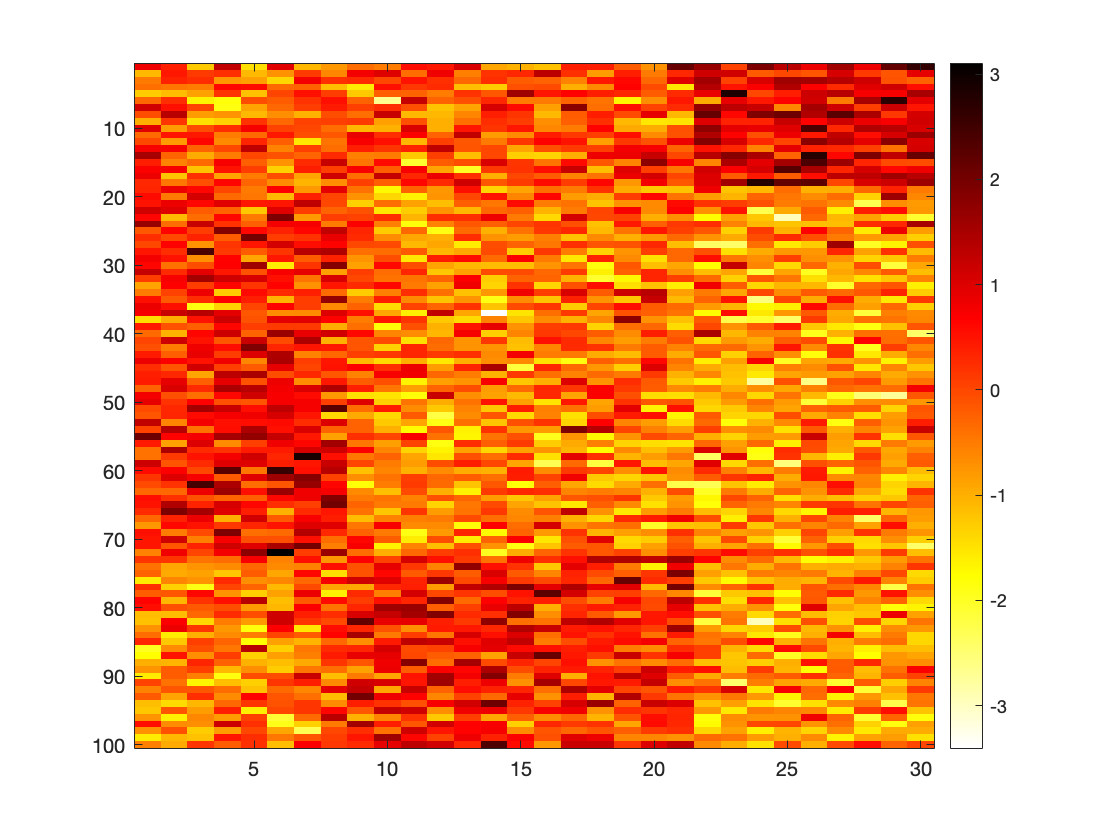}
        \caption{$\epsilon = 0.75$}
    \end{subfigure}
    \hfill
    \begin{subfigure}[b]{0.32\textwidth}
        \centering
    \includegraphics[width=\textwidth]{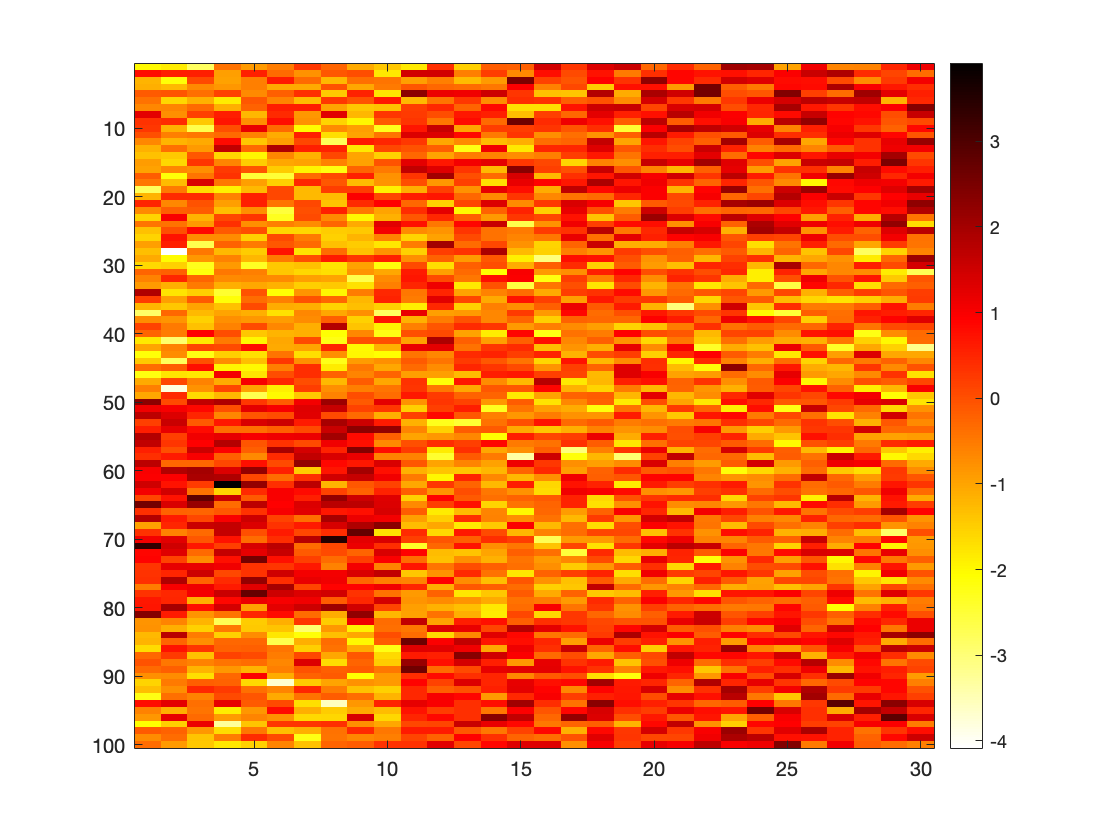}
        \caption{$\epsilon = 0.9$}
    \end{subfigure}
\end{figure}

For simplicity, we set both error terms to the same value in each simulation run (\( \epsilon_{\text{centroid}} = \epsilon_{\text{cluster}} = \epsilon \)), where \( \epsilon \) takes on values of 0.1, 0.35, 0.5, 0.75, and 0.9.
The data at the different error levels are displayed in Figure~\ref{fig:heatmaps}.
By varying \( \epsilon \) simultaneously for both error sources, we assessed the combined impact of increasing noise levels on the ability of the pseudo-$F$ index to correctly identify the true number of clusters.

We performed 100 runs, varying K and Q from 2 to 5. For each run, we calculated the pseudo-$F$ index for each combination of K and Q and recorded the number of times each combination resulted in the highest pseudo-$F$ value.

The results, summarized in Table~\ref{tab:pseudoF_results_K4_Q3}, indicate that at a low error level ($\epsilon = 0.1$), the pseudo-$F$ index correctly identified the true number of clusters (\( K = 4, Q = 3 \)) in 100\% of the runs. However, as the error level increased, the pseudo-$F$ index increasingly favored solutions with fewer clusters. For instance, at \( \epsilon = 0.5 \), the correct combination was selected in only 13\% of the runs, while combinations with \( K = 2 \) or \( K = 3 \) and \( Q = 2 \) became more prevalent. At the highest error level (\( \epsilon = 0.9 \)), the pseudo-$F$ index did not select the true number of clusters in any run, instead favoring the combination \( (K = 2, Q = 2) \) in 70\% of the runs. 
These results indicate that the pseudo-$F$ index is robust in identifying the correct number of clusters in data with a clear cluster structure but tends to underestimate the number of clusters when the data become noisier.

\begin{table}[H]
\centering
\caption{Number of times the true combination \((K=4, Q=3)\) had the highest pseudo-$F$ index across 100 runs for different error levels \( \epsilon \).}
\label{tab:pseudoF_results_K4_Q3}
\begin{tabular}{@{}llcccc@{}}
\toprule
& & \multicolumn{4}{c}{\( Q \)} \\
\cmidrule(lr){3-6}
\( \epsilon \) & \( K \) & 2 & 3 & 4 & 5 \\
\midrule
\multirow{4}{*}{0.1}  & 2 & 0   & 0   & 0 & 0 \\
                      & 3 & 0   & 0   & 0 & 0 \\
                      & 4 & 0   & \textbf{100} & 0 & 0 \\
                      & 5 & 0   & 0   & 0 & 0 \\
\midrule
\multirow{4}{*}{0.35} & 2 & 0   & 0   & 0 & 0 \\
                      & 3 & 15  & 0   & 0 & 0 \\
                      & 4 & 0   & \textbf{85}  & 0 & 0 \\
                      & 5 & 0   & 0   & 0 & 0 \\
\midrule
\multirow{4}{*}{0.5}  & 2 & 20  & 0   & 0 & 0 \\
                      & 3 & 66  & 0   & 0 & 0 \\
                      & 4 & 1   & \textbf{13}  & 0 & 0 \\
                      & 5 & 0   & 0   & 0 & 0 \\
\midrule
\multirow{4}{*}{0.75} & 2 & 58  & 1   & 0 & 0 \\
                      & 3 & 37  & 2   & 0 & 0 \\
                      & 4 & 0   & \textbf{2}   & 0 & 0 \\
                      & 5 & 0   & 0   & 0 & 0 \\
\midrule
\multirow{4}{*}{0.9}  & 2 & 70  & 0   & 0 & 0 \\
                      & 3 & 30  & 0   & 0 & 0 \\
                      & 4 & 0   & \textbf{0}   & 0 & 0 \\
                      & 5 & 0   & 0   & 0 & 0 \\
\bottomrule
\end{tabular}
\end{table}

It must be noticed that performing 100 simulation runs on a MacBook equipped with an Apple M1 chip and 8 GB of RAM, took on average approximately 87 seconds for the total running time, meaning each run took less than a second on average. This indicates a substantial reduction in computation time compared to traditional SKM algorithms, which are typically computationally intensive \citep{DUWAIRI201549, KIM2020113288}. The improved efficiency is attributed to our use of matrix operations in the implementation, which streamlines calculations and minimizes processing overhead.

\subsection{Selection of Random Starts}
Since the partitioning problem is known to be NP-hard, the new methodology is not guaranteed to obtain a global solution.
Therefore, this Section is devoted to discuss the selection of Random Starts (\textit{RndStarts}).
\par First, model \ref{data_gen} is used to generate a data matrix $\bf{X}$ with a high error level (\( \epsilon_{\text{centroid}} = \epsilon_{\text{cluster}} = \epsilon = 1.5 \)): therefore, the true partitions $\bf{U}$, $\bf{V}$, the true centroids $\bf{\overline{Y}}$ and the true value of the objective function are known.
\par Then, the algorithm is run 100 times by letting the number of \textit{RndStarts} take values in [1, 5, 20, 30, 40, 50, 70, 100].
For each \textit{RndStart} the algorithm is therefore run 100 times and the values of the objective function are stored. At the end, for each \textit{RndStart} we obtain a distribution of $100$ values of the objective function. Clearly, it is possible to compare those values with the true one. Whenever the algorithm returns a value of the objective function lower than the true one, the algorithm is trapped in a local maximum.
\par It is advised to select the number of random starts in such a way a local maximum never occurs. 
\par The percentage of local maxima occurrences for each random start are reported in Table~\ref{tab:local maxima}. The percentage of local maxima always decreases with the increase of the RndStarts until
reaching 0, when the number of random starts is set equal to 20. Thus, the selected
number of random starts for the whole simulation study was set to RndStarts = 20.
\begin{table}[h]
    \centering
        \caption{Local maxima occurrences (\%) with high error}
    \label{tab:local maxima}
    \begin{tabular}{c|c|c|c|c|c|c|c|c|c|}
    \toprule 
    Random Start & 1 & 5 & 10 & 20 & 30 & 40 & 50 & 70 & 100\\
         $\%$ of local maxima &  71 & 23 & 9 & 0 & 0 & 0 & 0 & 0 & 0 \\
    \bottomrule
    \end{tabular}

\end{table}
\newline From Table~\ref{tab:local maxima} it can be observed that with 1 random start of the algorithm and high error, the new methodology performs bad,
identifying the optimal solution in only $29\%$ of cases. It is worth noticing that with 10 random starts, the
algorithm identifies the global optimal solution in 91\% of cases, and with 20 random starts the algorithm was never trapped in local
maxima.

\subsection{Performance in terms of true partitions and centroids recovery}
We conducted a simulation study to evaluate the effectiveness of the algorithm in recovering the correct true partitions $\bf{U}$ and $\bf{V}$ and centroids matrix $\bf{\overline{Y}}$ under varying levels of error. 
In our simulations, we generated synthetic data with known cluster structures by using the model \ref{data_gen}. The data matrix \(\bm{X}\) was constructed based on the true cluster configurations, with \( K_{\text{true}} = 3 \) clusters for units and \( Q_{\text{true}} = 2 \) clusters for variables. 

For simplicity, we set both error terms to the same value in each simulation run (\( \epsilon_{\text{centroid}} = \epsilon_{\text{cluster}} = \epsilon \)), where \( \epsilon \) takes on values of 0.1, 0.35, 0.5, 0.75, 0.9, 1.1, 1.35, 1.5, 1.75, and 2. By varying \( \epsilon \) simultaneously for both error sources, we assessed the combined impact of increasing noise levels on the ability of the algorithm to recover the underlying partitions and the centroids matrix.

For each error level, we performed 500 runs, for a total of 5000 samples. For each run, and for each error level, we calculated the ARI index for both $\bf{U}$ and $\bf{V}$ and the RMSE and its normalized versions, NRMSE1 and NRMSE2. Obviously, the higher the ARI the better the similarity between the obtained and true partitions; the lower the RMSE, NRMSE1, NRMSE2, the closer the obtained and the true centroids matrices. The normalized versions of RMSE ranges is $[0,1]$.

The results, summarized in Table~\ref{tab:sim performance}, indicate that when the error level increases, the performance get worse, as expected. More in details, at a low error level ($\epsilon = 0.1$), the true partitions are perfectly recovered, as well as the centroid matrix. By increasing the error level, the ARI indices decreases and the RMSE, NRMSE1, NRMSE2 indices increase. For instance, with error level $\epsilon$ equal to 1.35, the summary statistics of ARI indices are about 0.7 and the average NRMSEs are close to 0.08. It is worth noticing that the median value of ARI for $\bf{V}$ remains equal to 1 by increasing the error level until 1.35, meaning that half of the runs completely recovers the true partition. Instead, when the error level reaches 1.75, the median ARI for $\bf{V}$ rapidly declines toward 0.522, meaning that in half of the runs, the true partition is recovered only partially.
When the error level is set to 2, then the partitions are not recovered: the obtained ARI indices are slightly above 0, value that is reached only when the comparing partition is a random one \citep{hubert1985comparing}. For what concerns the centroids matrix recovery, NRMSEs statistics are all above 0.1.

\begin{table}[ht]
    \centering
    \caption{Summary statistics to evaluate algorithm's performance under low and high level of error.}
    \label{tab:sim performance}
    \begin{tabular}{llcccccccccc}
    \toprule
        & & \multicolumn{10}{c}{Error level $\epsilon$}  \\
\cmidrule(lr){3-12}
index & statistic & 0.10 & 0.35 & 0.50 & 0.75 & 0.90 & 1.10 & 1.35 & 1.50 & 1.75 & 2.00 \\
\midrule
\multirow{2}{*}{ARI for U} & mean & 1 & 1 & 1 & 0.974 & 0.907 & 0.794 & 0.675 & 0.587 & 0.480 & 0.354\\
                          & median  & 1 & 1 & 1 & 1  & 0.976 & 0.877 & 0.682 & 0.567 & 0.467 & 0.360\\

\multirow{2}{*}{ARI for V} & mean & 1 & 1 & 1 & 0.984  & 0.933 & 0.840 & 0.717 & 0.649 & 0.507 & 0.332 \\
                           & median &  1 & 1 & 1  &1 & 1 & 1 & 1 & 0.866 & 0.522 & 0.192\\
\multirow{2}{*}{RMSE} & mean & 0.003 & 0.010 & 0.015 & 0.03 & 0.044&  0.082 & 0.132 & 0.167 & 0.225 & 0.314\\
                      & median & 0.003 & 0.009 & 0.015 & 0.02 & 0.031 & 0.047 & 0.070 & 0.098 & 0.159 & 0.260\\
\multirow{2}{*}{NRMSE1} & mean & 0.001 & 0.005 & 0.008 & 0.016 & 0.025 & 0.045 & 0.074 & 0.094 & 0.129 & 0.180\\
                        & median & 0.001 & 0.005 & 0.008 & 0.012 & 0.017 & 0.025 & 0.040 & 0.056 & 0.088 & 0.145\\
\multirow{2}{*}{NRMSE2} & mean & 0.002 & 0.006 & 0.009 & 0.018 & 0.027 & 0.050 & 0.082 & 0.104 & 0.143 & 0.198\\
                        & median & 0.002 & 0.006 & 0.008 & 0.013 & 0.018 & 0.027 & 0.043 & 0.061 & 0.099 & 0.162\\
\bottomrule

    \end{tabular}
    
\end{table}
To summarize, we can conclude that with an increasing level of error, the recovery of the true partitions become a really hard task, since the error terms mask the true underlying partition. On the contrary, with an increasing level of error, the recovery of the centroid matrix is still quite feasible, as the summary values of the NRMSEs never reach or become close to their maximum, i.e. 1. 

\section{Application}
\label{sec:application}

We applied the SDKM method to a dataset containing U.S. presidential inaugural addresses, ranging from those delivered by George Washington in 1789 to Joe Biden in 2021. Our objective is to identify shared topics among speeches given by different presidents. Using SDKM allows for the simultaneous clustering of both documents and terms, providing a comprehensive view of the evolving presidential discourse. 
The initial corpus comprises 59 documents, with 151,536 tokens, 9,442 types, and 4,218 hapaxes. On average, each document contains 770.4 types and 2,568 tokens. We calculated the type/token ratio, 0.0623, and the hapax percentage, 0.4467. These values suggest a high lexical variety, justifying the computational analysis of the corpus. 

In our analysis, data cleaning and preprocessing were crucial. This involved tokenizing texts and removing extraneous elements such as numbers and punctuation. Lemmatization was applied to reduce each word to its base form. We built a document-term matrix, excluding stop-words and trimming infrequent words with a frequency of 11 or less. These steps were necessary to focus the analysis on core topics articulated by the presidents. We employed the TF-IDF weighting scheme, which assesses a word’s significance within documents relative to the entire corpus (\citep{liang22}, \cite{manning2009introduction}). This approach helps mitigate length-induced bias in document analysis by focusing on term importance rather than sheer frequency. 
 Specifically, TF-IDF measures how important a term is within a document relative to a corpus, and it is calculated as: 
\begin{equation*}
    x_{ij}=\frac{n_{ij}}{n_{\cdot j}}\log_{10}\frac{M}{m},
\end{equation*}
where $n_{ij}$ is the raw number of occurrences, $M$ is the number of documents and $m$ is the number of documents that include the term. 
The final data matrix, comprising 59 documents and 1,206 types, demonstrated a sparsity of 67.73\%

\subsection{Spherical Double k-means}

To determine the best number of clusters for both terms and documents, we used the Pseudo-$F$ index with K and Q ranging from 2 to 10. Table~\ref{tab:pseudoF} presents these Pseudo-F values, highlighting the clustering dynamics.

\begin{table}[H]
\centering
\caption{Pseudo-F values for combinations of $K$ and $Q$.}
\label{tab:pseudoF}
\begin{tabular}{c|ccccccccc}
\hline
$K \backslash Q$ & 2 & 3 & 4 & 5 & 6 & 7 & 8 & 9 & 10 \\
\hline
2 & \textbf{1199.5} & 675.7 & 655.6 & 410.5 & 452.4 & 236.1 & 236.2 & 286.9 & 212.9 \\
3 & \textbf{707.3}  & 593.9 & 351.9 & 241.8 & 268.8 & 190.3 & 162.9 & 183.6 & 181.7 \\
4 & 536.6  & 417.9 & 344.4 & 278.1 & 167.5 & 168.0 & 140.4 & 131.2 & 103.2 \\
5 & 439.7  & 294.0 & 258.0 & 166.1 & 187.6 & 130.3 & 91.7  & 88.7  & 131.2 \\
6 & 358.8  & 257.0 & 193.4 & 150.4 & 153.7 & 142.2 & 99.7  & 62.9  & 75.5  \\
7 & 326.8  & 190.5 & 172.1 & 134.7 & 90.8  & 91.8  & 91.0  & 59.2  & 57.1  \\
8 & 273.3  & 192.0 & 162.5 & 126.2 & 139.1 & 84.1  & 76.2  & 76.5  & 67.6  \\
9 & 247.0  & 167.4 & 116.4 & 107.3 & 91.6  & 73.3  & 64.1  & 64.6  & 50.8  \\
10 & 208.5  & 165.2 & 127.0 & 98.1  & 82.1  & 62.8  & 70.4  & 46.0  & 50.8  \\
\hline
\end{tabular}
\end{table}

\begin{figure}[H]
\caption{Clustergram of $10\times 10$ centroids matrix.}
    \label{fig:clustergram}
    \centering
    \includegraphics[width=0.5\linewidth]{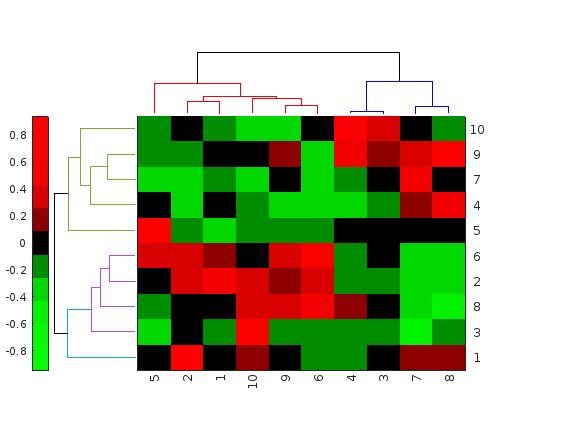}
\end{figure}

\begin{figure}[H]
\caption{Clustergram of the theoretical data matrix narrowed down to 150 sampled rows.}
    \label{fig:clustergram_sample}
    \centering
    \includegraphics[width=0.5\linewidth]{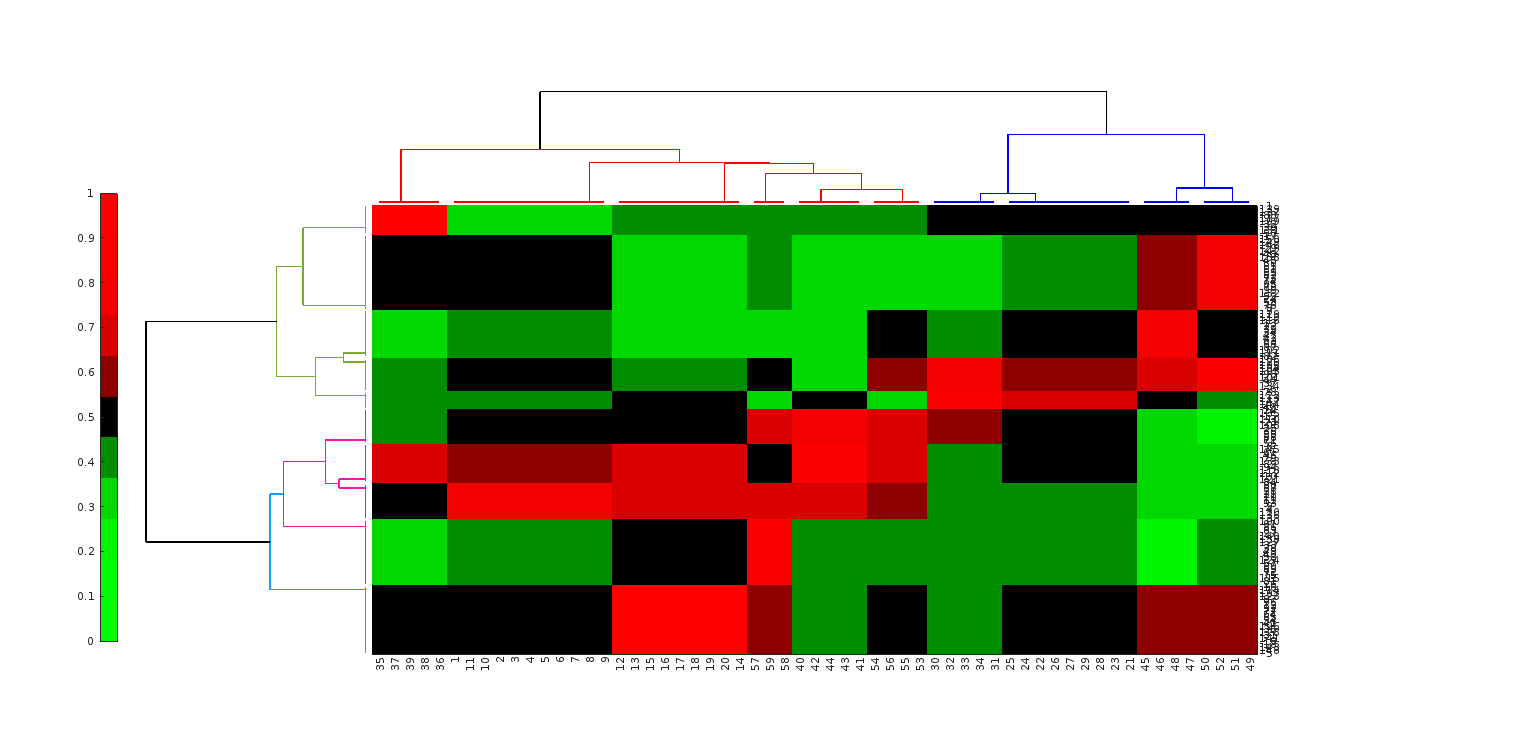}
\end{figure}

The highest Pseudo-$F$ value occurs at $K=2$ and $Q=2$, suggesting that the optimal number of clusters for both words and documents is 2. However, as discussed in Section~\ref{subsec: pseudo-F simulation}, the Pseudo-$F$ index can tend to underestimate the number of clusters. The second highest Pseudo-$F$ value occurs at $K=3$ and $Q=2$. The observed patterns in the clusters indicate that the choice of $K$ = 3 and $Q$ = 2 produces clusters with greater interpretability, a crucial factor in the effective analysis of clusters as fully discussed in \citep{FraleyHowManyClusters}. In addition, to further investigate this aspect, we apply our methodology with $K=10$ and $Q=10$ and we plot the clustergram of the centroid matrix. The clustergram in Figure~\ref{fig:clustergram}, which shows the hierarchical clustering of the rows and columns of the centroids matrix through a heatmap and two dendrograms, reveals $K=3$ distinct clusters of words and $Q=2$ distinct clusters of documents. The dendrograms are obtained by using one minus the sample correlation between points as distance metric for both rows and columns and single linkage and Ward linkage as hierarchical clustering methods, for rows and columns, respectively. In addition, the same clustergram procedure is applied to a sample of $150$ rows of the theoretical data matrix reconstructed via: $\mathbf{X}_t=\mathbf{U}\bar{\mathbf{Y}}\mathbf{V}^\prime$. Figure~\ref{fig:clustergram_sample} shows again $K=3$ distinct clusters of words and $Q=2$ distinct clusters of documents.  
Therefore, $K$ = 3 and $Q$ = 2  was our final choice.

The analysis provides insights into the evolution of themes in U.S. presidential inaugural addresses.
The three clusters of words contain 457 (37.9\%), 420 (34.9\%) and 328 (27.2\%) words, while the two groups of documents contain 31 and 28 documents.

\subsubsection{Document Distribution and Historical Context}

Figure~\ref{DOCclust} presents the document distribution across SDKM’s two clusters, along with key events in U.S.\ history. Cluster 1 spans speeches from 1789 to 1861, reflecting the early nation-building era, while Cluster 2 covers addresses from 1865 onward, aligning with the modern evolution of presidential rhetoric.

\begin{center}
\captionof{figure}{Historical timeline of the US presidents, colored by cluster of documents and their 5 most frequent words (TF-IDF normalized).}
    \label{DOCclust}
    \begin{subfigure}{\textwidth}
        \centering
        \includegraphics[width=.9\textwidth]{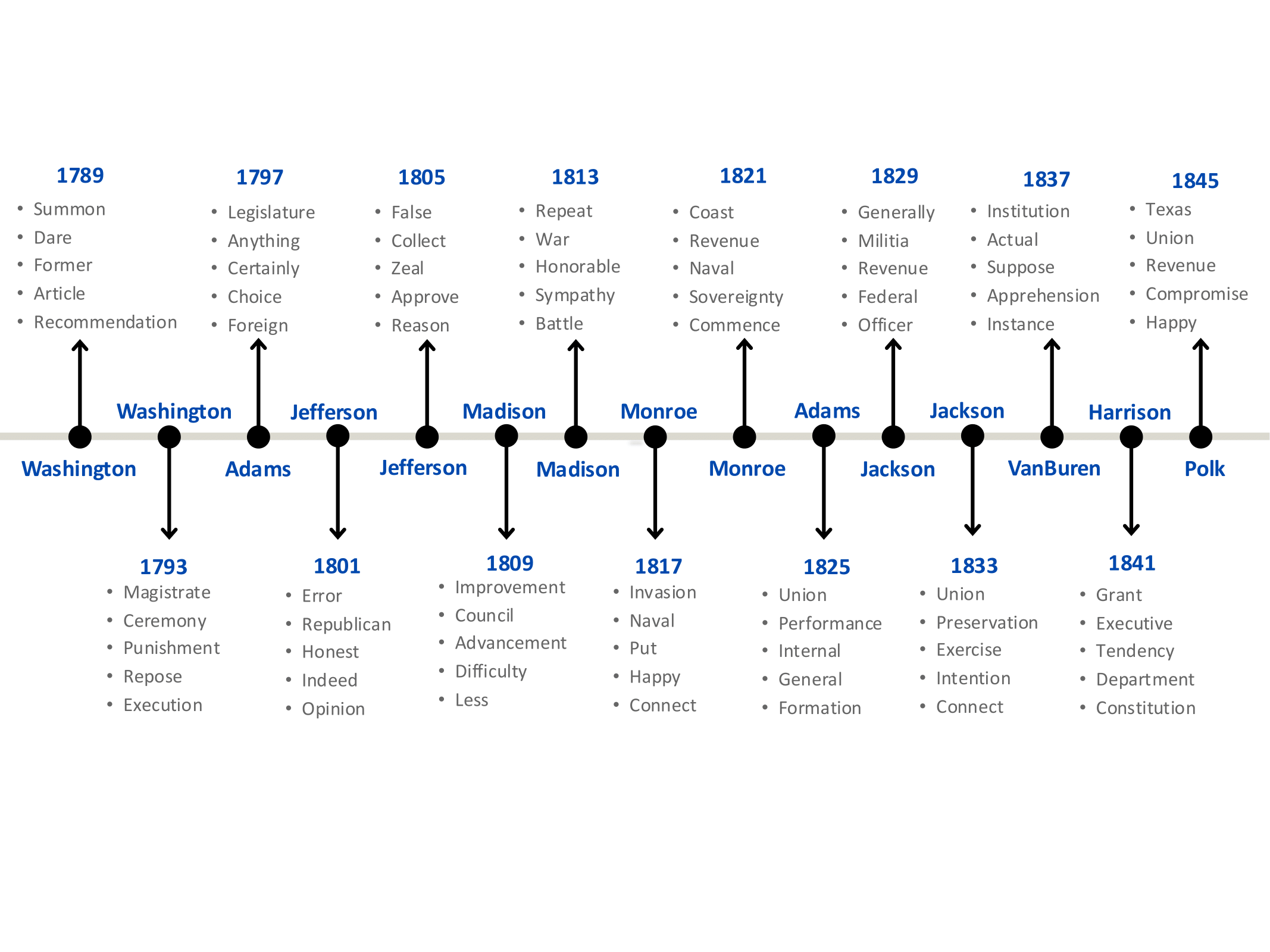}
    \end{subfigure}
    
    \vspace{-8mm}
    
    \begin{subfigure}{\textwidth}
        \centering
        \includegraphics[width=.9\textwidth]{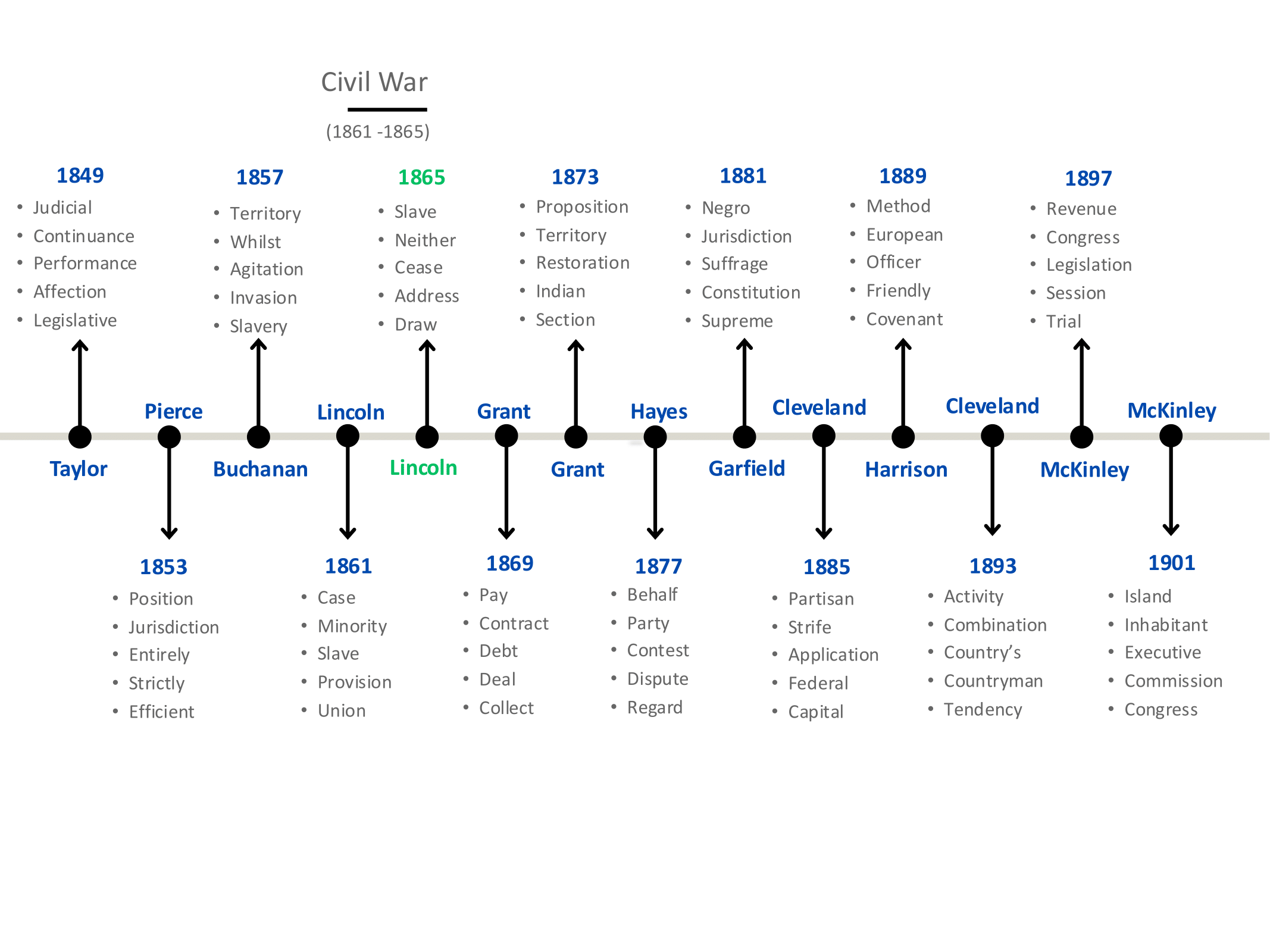}
    \end{subfigure}
    
    \vspace{-8mm}
    
    \begin{subfigure}{\textwidth}
        \centering
        \includegraphics[width=.9\textwidth]{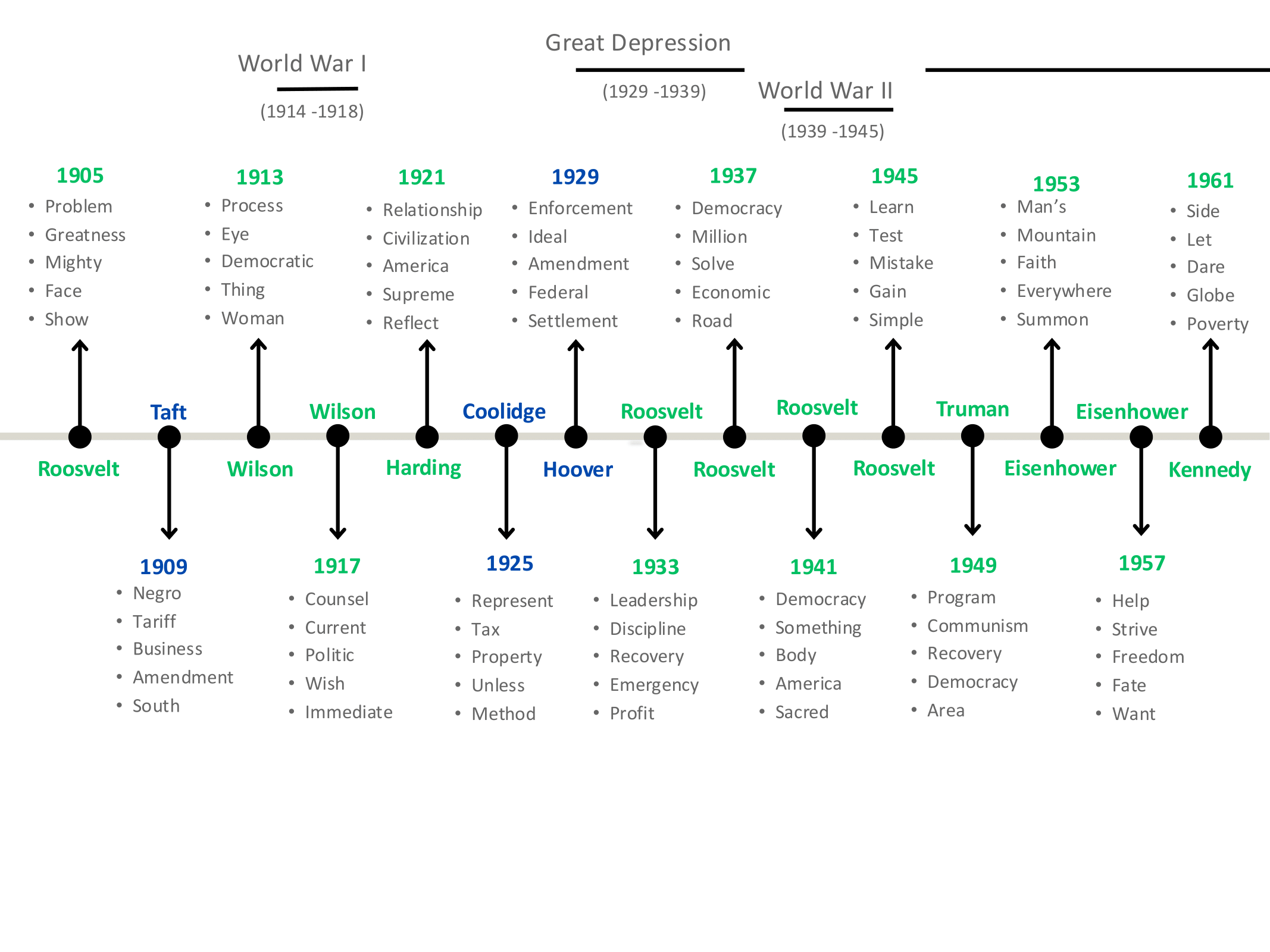}
    \end{subfigure}
    
    \vspace{-8mm}
    
    \begin{subfigure}{\textwidth}
        \centering
        \includegraphics[width=.9\textwidth]{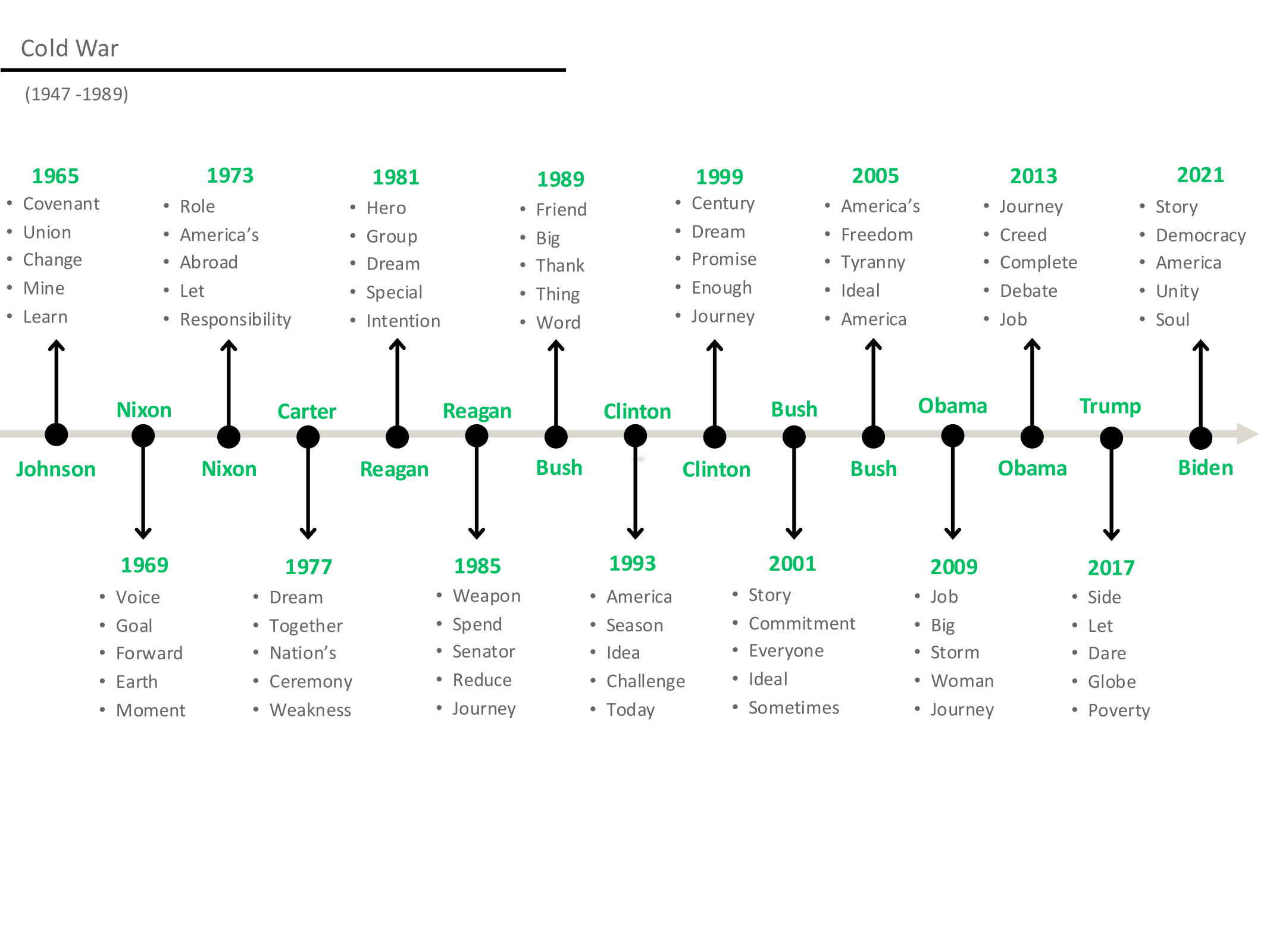}
    \end{subfigure}
\end{center}

The division between clusters aligns closely with significant historical periods.  Lincoln's 1861 speech, placed in cluster 1, focuses on the legal implications of secession and efforts to preserve the union. In contrast, his 1865 speech, included in cluster 2, emphasizes reconciliation and future aspirations, aligning with themes of national healing more typical of the recent years cluster. Interestingly, Calvin Coolidge (1925) and Herbert Hoover (1929) are later exceptions in Cluster 1. Their speeches likely focus on governmental and economic policies prior to the great depression, which aligns more closely with the themes prevalent in cluster 1.

\subsubsection{Word Clusters and Thematic Interpretation}

\begin{figure}[H]
\centering
\caption{Top 30 terms in each of the three word clusters (TF-IDF weighted).}
\label{WORDSclust}
\includegraphics[width=\linewidth]{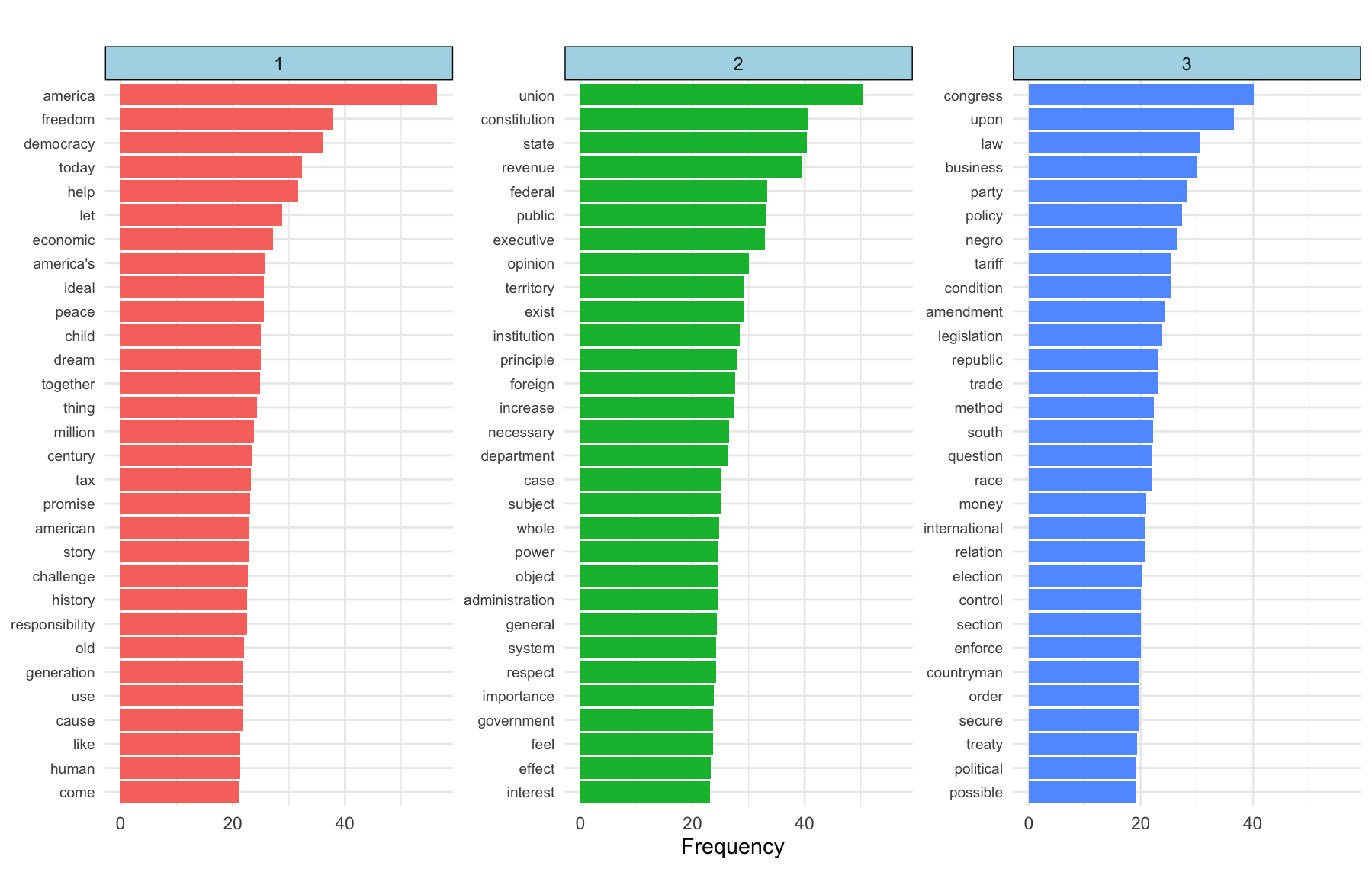}
\end{figure}

Figure~\ref{WORDSclust} displays the top 30 terms in each of the three word clusters. Cluster~1, that we labeled “\textbf{American Dream},” features words like ``america,'' ``freedom,'' ``democracy,'' and ``dream,'' pointing to aspirational rhetoric and national ideals. Cluster~2 has been labeled “\textbf{Law and Order},” it centers on terms like “union,” “constitution,” “state,” and “government,” underscoring legal and institutional frameworks. We called Cluster~3, “\textbf{Politics, Economics, and Secession}.” It includes words such as “congress,” “law,” “business,” and “policy,” reflecting deeper political processes and economic discussions.

\begin{figure}[H]
\caption{Each point represents a single inaugural address, colored by its document-cluster membership (blue for cluster~1, green for cluster~2). The coordinates are mean TF-IDF values of each address in the three word-clusters.}
\label{fig:documents_3D}
\centering
\includegraphics[width=\linewidth]{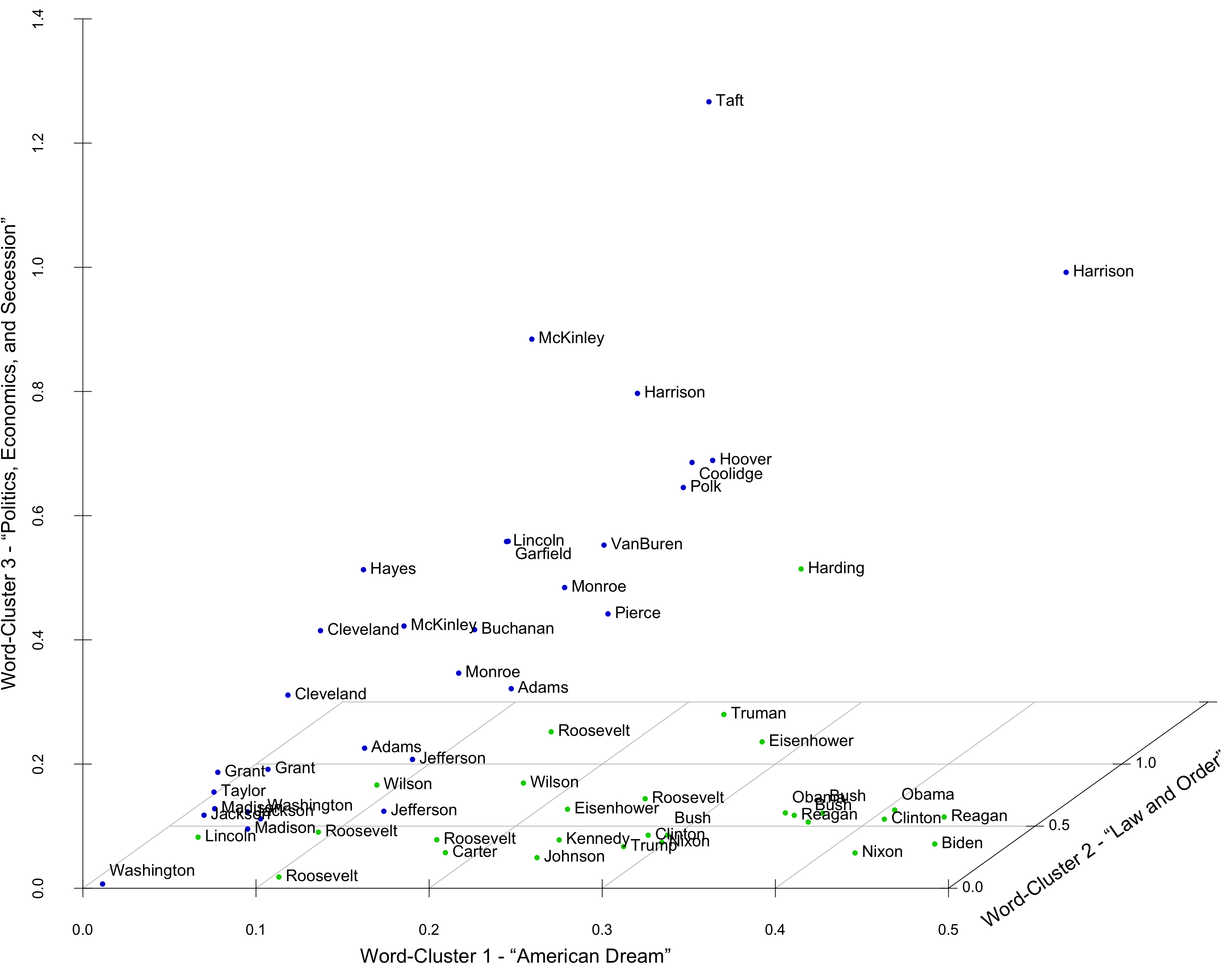}
\end{figure}

Figure~\ref{fig:documents_3D} further illustrates the document-level clustering in a pseudo-3D coordinate system, where each axis corresponds to a cluster of words. Notably, speeches in the “Politics, Economics, and Secession” dimension align more strongly with the older cluster, while documents with higher “American Dream” or “Law and Order” components fall into the more modern cluster. Addresses like Lincoln’s 1865 or Coolidge’s 1925 appear in positions consistent with their heavier emphasis on unity or policy-related concerns, linking them more closely to modern rhetorical changes.

\begin{figure}[H]
\centering
\caption{Most frequent words in each document cluster, colored by their respective word clusters.}
\label{DOCxWORDSclust}
\includegraphics[width=\linewidth]{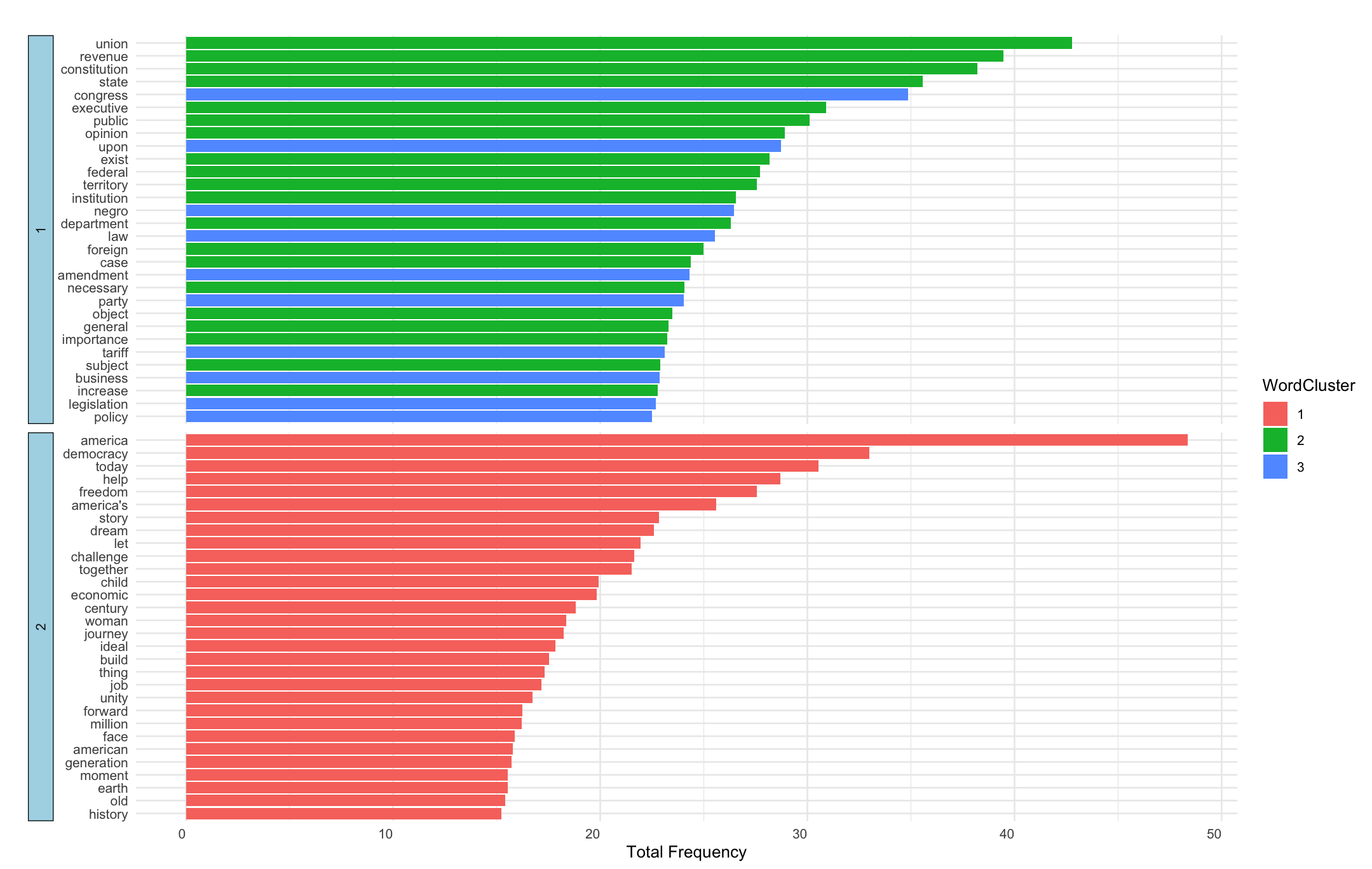}
\end{figure}

Finally, Figure~\ref{DOCxWORDSclust} examines the 30 most frequent words in each document cluster, colored by word-cluster membership. The second document cluster draws heavily on the “American Dream” word set, whereas the first cluster incorporates terms largely from “Law and Order,” as well as some from “Politics, Economics, and Secession.” This pattern aligns with the nation’s historical development: early speeches concerned with legal underpinnings and secession disputes, and later addresses emphasizing aspirational, forward-facing themes.

Altogether, these results highlight how the themes covered by the presidents' addresses evolved in time and how major historical events, such as the Civil War, world wars, and the Great Depression, shaped the language and focus of presidential inaugural addresses, illustrating a broader shift from foundational legal concerns to modern themes of national identity and aspirations for the future.

\section{Comparison of DKM and SDKM Clusters}
\label{sec:comparison}

We applied Double K-Means (DKM) on the same data (preprocessed and TF-IDF weighted) to compare the results with those achieved using SDKM. We examine both document-level and word-level clusters, highlighting how Euclidean distance (DKM) versus cosine similarity (SDKM) can produce subtle but meaningful differences in this double clustering setting.
We do not dwell on the comparison of computational times, as they are negligible in both methods.

\subsection{Document Clusters: Lincoln 1865 as the Only Difference}

Regarding the documents cluster membership, strikingly, every speech falls into the same two‐cluster partition across both methods, earlier historical presidents vs.\ later ones, except for Lincoln’s 1865 address.

\begin{itemize}
  \item \textbf{DKM:} Places Lincoln’s 1865 inaugural in the older (founding era) cluster. Even with TF-IDF weighting, Euclidean distance reflects the overall magnitude of the feature vector; Lincoln’s 1865 speech aligns more closely with earlier addresses in that sense.
  \item \textbf{SDKM:} Classifies Lincoln 1865 among more modern addresses. By focusing on relative usage proportions of words like “union,” “heal,” “mercy,” and “nation,” SDKM finds it more thematically aligned with 20th‐century rhetoric (rather than strictly early 19th‐century addresses).
\end{itemize}

\subsection{Word Clusters: Thematic Comparisons}

Beyond their broad similarity in document partitioning, DKM and SDKM exhibit noteworthy differences at the word level. To illustrate how each method groups the main words, we build a contingency table, represented in Table~\ref{tab:DKM_SDKM_contig}, that compares the cluster assignments for all words. Specifically, each row corresponds to a DKM cluster, and each column to a SDKM cluster.

\begin{table}[H]
\centering
\caption{Contingency table for all words assigned to each pair of clusters (DKM rows, SDKM columns).}
\label{tab:DKM_SDKM_contig}
\begin{tabular}{c|ccc}
\hline
\textbf{DKM $\backslash$ SDKM} & 1 & 2 & 3\\
\hline
1 & 201 & 0   & 0 \\
2 & 0   & 295 & 143 \\
3 & 256 & 125 & 185 \\
\hline
\end{tabular}
\end{table}

The Table reveals how words are distributed across the six possible cluster pairs. For example, 201 words appear in both cluster~1 under DKM and cluster~1 under SDKM, whereas 256 words assigned to DKM~3 fall into SDKM~1. These differences reflect how each method partitions the same vocabulary, highlighting areas of overlap and divergence. To further investigate differences in the two partitions, we use the ARI to measure the similarity between two partitions; it usually ranges in $[0,1]$ and it is equal to 1 when the two partitions perfectly match, while it is equal to 0 when one of the two partitions can be considered as a partition obtained by chance. The two partitions are dissimilar, as also the ARI index equal to 0.225 reveals. 

By exploring differences in the clustering interpretation, we considered the ''top words'' (words with the 30 highest frequencies) in each cluster. In fact, these words help to label each cluster with a specific topic.

\begin{figure}[H]
\caption{Top words in each cluster for DKM vs.\ SDKM. Purple bars indicate overlap in both methods, green indicates DKM only, and orange indicates SDKM only.}
\label{fig:topwords_comparison}
\centering
\includegraphics[width=\linewidth]{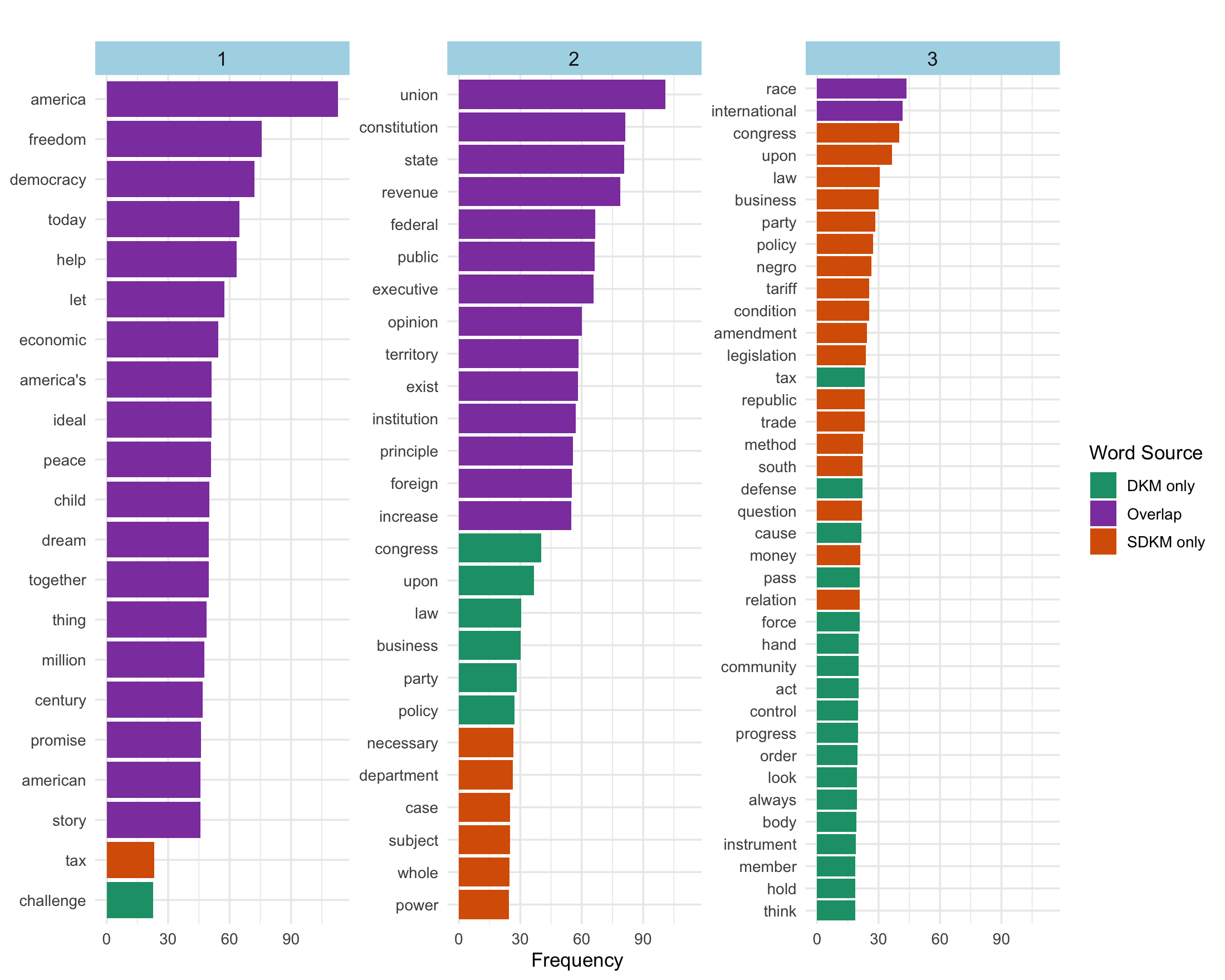} 
\end{figure}

Figure~\ref{fig:topwords_comparison} provides a complementary view of these differences. In each panel, we compare the top words identified by DKM and by SDKM for the same cluster index. If a word appears in both methods’ top~30, it is shown in purple (``Overlap''), while words unique to DKM or SDKM are colored differently. 

Both DKM and SDKM isolate a cluster of aspirational or national-identity terms, including \emph{“america,” “freedom,” “democracy,” “child,” “peace,” and “dream.”} These words typically co-occur in modern addresses (post–Civil War and throughout the 20th century), emphasizing a forward-looking vision for the nation. Because such vocabulary dominates many speeches from that era, both Euclidean- and cosine-based approaches identify a similar \emph{``American Dream''} cluster, which is apparent in Figure~\ref{fig:topwords_comparison} through the substantial overlap (purple bars) in the first panel.
The second cluster, highlighted by words like \emph{“union,” “constitution,” “state,” and “revenue,”} centers on governance and legal structures. These foundational words appear mostly as overlap, reflecting how both methods capture a \emph{“Law and Order”} theme. Nevertheless, DKM alone includes terms such as \emph{“congress”} and \emph{“law”} in its top list, whereas SDKM alone brings in additional administrative or procedural vocabulary such as \emph{“case”} and \emph{“department.”} This suggests that SDKM’s emphasis on relative usage encourages a slightly finer partitioning of specialized policy references, whereas DKM merges them when they appear together in large documents.
The third cluster, focusing on politics, economic considerations, and older secession-related rhetoric, shows somewhat greater divergence. Figure~\ref{fig:topwords_comparison} reveals that SDKM uniquely elevates words like \emph{“negro,” “tariff,”} and \emph{“amendment,”} while DKM includes \emph{“community,” “control,”} and \emph{“progress”} in the same top set. Although certain vocabulary, such as \emph{“race”} and \emph{“congress,”} again appears in both methods’ top words, the larger number of unique terms indicates that SDKM splits off certain policy and historical references more decisively, presumably by tracking proportional usage across speeches. DKM, on the other hand, merges them whenever they tend to co-occur in lengthier addresses, thereby placing broad economic and societal concepts together.
These patterns underscore how SDKM’s cosine-based metric isolates specific rhetorical proportions, creating distinct clusters for governance/constitutional issues versus modern policy/economic themes. Meanwhile, DKM groups some topics together whenever they coexist in the same documents, downplaying subtle proportional distinctions. This aligns with the document-level discrepancy observed in Lincoln’s 1865 speech: a borderline case, rich in ``union'' and ``heal'' language, can shift clusters under SDKM, just as certain policy or war-related terms shift between the two methods depending on how they recur in the corpus.

\section{Final considerations}
\label{sec:conclusion}

In this paper, we introduced the Spherical Double K-Means clustering method, a novel approach for the simultaneous partitioning of terms and documents in textual data. By integrating the principles of K-Means, Double K-Means, and Spherical K-Means clustering, SDKM effectively addresses the challenges posed by high dimensionality, sparsity, and noise inherent in text data. The incorporation of cosine similarity within the co-clustering framework allows for a deep understanding of the text and a more meaningful clustering by considering the angular relationships between data points, which is particularly beneficial for textual data where document lengths and term frequencies can vary significantly.
In fact, traditional k-means using Euclidean distance often produces large, dominant clusters alongside smaller residual ones when applied to textual data, even after TF-IDF weighting. This tendency occurs because Euclidean distance is most effective when variables are uncorrelated, a condition seldom met in language data where terms frequently co-occur in highly correlated patterns. By contrast, SDKM, employing cosine similarity, focuses on the relative distribution of term usage rather than absolute frequencies. In doing so, it captures the angular relationships between document vectors, thereby disentangling topics more effectively and yielding clusters that are more semantically coherent.
Applying SDKM to the US presidential inaugural addresses dataset demonstrated its capability to uncover distinct thematic clusters that evolved over time. The method not only identified clear chronological distinctions in the clusters of documents but also revealed how historical events have influenced the themes and language used in presidential discourse. For instance, the transition between clusters aligns with significant periods such as the civil war, the great depression and world wars, highlighting shifts in national priorities and rhetoric. This application underscores the potential of SDKM to extract meaningful patterns from complex text corpora, making it a valuable tool for tasks such as topic modeling, sentiment analysis, and information retrieval.

In addition to demonstrating SDKM's effectiveness on the presidential addresses corpus, we compared its results with Double K-Means using the same TF-IDF weighted data. While both methods identified a broadly consistent chronological split for most documents, one notable discrepancy emerged with Lincoln’s 1865 speech, which SDKM grouped among more modern addresses. 
At the word level, the methods agreed on an “American Dream” cluster but diverged significantly on governance and policy related terms, indicating that SDKM’s cosine‐based co-clustering can isolate finer proportional distinctions. These findings reinforce SDKM’s potential to uncover nuanced patterns in complex textual corpora, complementing more traditional Euclidean-based approaches.

Despite the effectiveness of SDKM, there are avenues for further improvement and research. One area of future development is the implementation of a fuzzy version of SDKM. The current method assigns each term and document exclusively to one cluster, which may not fully capture the nuanced relationships in text data where terms and documents can naturally belong to multiple topics or themes. A fuzzy clustering approach would allow for degrees of membership, providing a more flexible and realistic representation of the data's inherent structure.

An open issue relates to the normalization of the centroid matrix within the co-clustering framework. Ideally, to achieve spherical clusters in both dimensions, the centroid matrix should be normalized by both rows and columns. However, simultaneous normalization poses challenges due to the interdependence of term and document clusters in co-clustering. In our approach, we opted to normalize the centroid matrix by rows, ensuring that the term vectors (rows) have unit length. This decision was based on the consideration that terms typically exhibit more variability and noise than documents. By focusing on the sphericality of term clusters, we enhance the method's robustness against noise in the data. While the norms of the columns (documents) in the centroid matrix were not explicitly normalized, they were observed to be close to one, suggesting that the lack of column normalization may not significantly impact the clustering results. Future work could explore alternative normalization strategies or iterative normalization procedures to address this issue more comprehensively.

\section*{Acknowledgements}

The views and opinions expressed are those of the authors and do not necessarily reflect the official policy or position of the Italian National Institute of Statistics - Istat.

\bibliography{MAIN_NEW}

\appendix

\section{Monotonicity Proof for SDKM}
\label{appendix:demonstrations}

This appendix demonstrates that the SDKM objective function does not decrease at each iteration, ensuring convergence to a local maximum. Specifically, if 
\(\bigl(\bm{U}^{(t)}, \bm{\overline{Y}}^{(t)}, \bm{V}^{(t)}\bigr)\) denotes the solution at iteration \(t\), we want to show
\[
  f\bigl(\bm{U}^{(t)}, \bm{\overline{Y}}^{(t)}, \bm{V}^{(t)}\bigr) 
  \;\;\le\;\;
  f\bigl(\bm{U}^{(t+1)}, \bm{\overline{Y}}^{(t+1)}, \bm{V}^{(t+1)}\bigr),
\]
where
\[
  f(\bm{U}, \bm{\overline{Y}}, \bm{V}) 
  \;=\; 
  \text{tr}\bigl(\bm{X}' \,\bm{U}\,\bm{\overline{Y}}\,\bm{V}'\bigr).
\]
By proving that each sub-step (updating \(\bm{U}\), then \(\bm{\overline{Y}}\), then \(\bm{V}\), then \(\bm{\overline{Y}}\) again) does not reduce \(f\), we establish that the full iteration is \emph{monotonically non-decreasing}.

With \(\pi_k^{(t)}\) is denoted the set of row observations (units) assigned to cluster \(k\) under \(\bm{U}^{(t)}\).
\(\tau_q^{(t)}\) denotes the set of column features assigned to cluster \(q\) under \(\bm{V}^{(t)}\).
Each row \(\bm{\overline{x}}_k\) in \(\bm{\overline{Y}}\) is viewed as the row-centroid for cluster \(k\).

At iteration \(t\), SDKM updates:
\[
(\bm{U}^{(t)}, \bm{\overline{Y}}^{(t)}, \bm{V}^{(t)}) 
\;\to\; 
(\bm{U}^{(t+1)}, \bm{\overline{Y}}^{(t+1)}, \bm{V}^{(t+1)})
\]
by successively maximizing the objective with respect to \(\bm{U}\), then \(\bm{\overline{Y}}\), then \(\bm{V}\), and finally \(\bm{\overline{Y}}\) again.

\noindent
\textbf{Step 1: Updating \(\bm{U}\).} 

\[
  f\bigl(\bm{U}^{(t)}, \bm{\overline{Y}}^{(t)}, \bm{V}^{(t)}\bigr)
  \;=\;
  \text{tr}\bigl(\bm{X}'\,\bm{U}^{(t)}\,\bm{\overline{Y}}^{(t)}\,\bm{V}^{(t)}\bigr).
\]
Because \(\bm{U}^{(t+1)}\) is chosen to maximize the objective for fixed \(\bm{\overline{Y}}^{(t)}\) and \(\bm{V}^{(t)}\), we have
\begin{equation}
  f\bigl(\bm{U}^{(t)}, \bm{\overline{Y}}^{(t)}, \bm{V}^{(t)}\bigr)
  \;\;\le\;\;
  f\bigl(\bm{U}^{(t+1)}, \bm{\overline{Y}}^{(t)}, \bm{V}^{(t)}\bigr).
\label{eq:Uupdate}
\end{equation}

\medskip
\noindent
\textbf{Step 2: Updating \(\bm{\overline{Y}}\).}

Holding \(\bm{U}^{(t+1)}\) and \(\bm{V}^{(t)}\) fixed, we update \(\bm{\overline{Y}}\) to \(\bm{\overline{Y}}^{(t+1)}\) by maximizing \(\text{tr}(\bm{X}' \,\bm{U}^{(t+1)}\,\bm{\overline{Y}}\,\bm{V}^{(t)})\). Consequently,
\begin{equation}
  f\bigl(\bm{U}^{(t+1)}, \bm{\overline{Y}}^{(t)}, \bm{V}^{(t)}\bigr)
  \;\;\le\;\;
  f\bigl(\bm{U}^{(t+1)}, \bm{\overline{Y}}^{(t+1)}, \bm{V}^{(t)}\bigr).
\label{eq:Yupdate1}
\end{equation}

Combining \eqref{eq:Uupdate} and \eqref{eq:Yupdate1} yields
\[
  f\bigl(\bm{U}^{(t)}, \bm{\overline{Y}}^{(t)}, \bm{V}^{(t)}\bigr)
  \;\le\;
  f\bigl(\bm{U}^{(t+1)}, \bm{\overline{Y}}^{(t+1)}, \bm{V}^{(t)}\bigr).
\]

\medskip
\noindent
\textbf{Step 3: Updating \(\bm{V}\).}

Now, for fixed \(\bm{U}^{(t+1)}\) and \(\bm{\overline{Y}}^{(t+1)}\), updating \(\bm{V}^{(t)}\to \bm{V}^{(t+1)}\) once again does not lower the objective:
\begin{equation}
  f\bigl(\bm{U}^{(t+1)}, \bm{\overline{Y}}^{(t+1)}, \bm{V}^{(t)}\bigr)
  \;\;\le\;\;
  f\bigl(\bm{U}^{(t+1)}, \bm{\overline{Y}}^{(t+1)}, \bm{V}^{(t+1)}\bigr).
\label{eq:Vupdate}
\end{equation}

\medskip
\noindent
\textbf{Step 4: Final Re-update of \(\bm{\overline{Y}}\).}

After \(\bm{V}\) is updated, the algorithm updates the centroids \(\bm{\overline{Y}}^{(t+1)}\) to \(\bm{\overline{Y}}^{(t+2)}\). Again, by maximizing \(\text{tr}(\bm{X}' \,\bm{U}^{(t+1)}\,\bm{\overline{Y}}\,\bm{V}^{(t+1)})\), we have
\begin{equation}
  f\bigl(\bm{U}^{(t+1)}, \bm{\overline{Y}}^{(t+1)}, \bm{V}^{(t+1)}\bigr)
  \;\;\le\;\;
  f\bigl(\bm{U}^{(t+1)}, \bm{\overline{Y}}^{(t+2)}, \bm{V}^{(t+1)}\bigr).
\label{eq:Yupdate2}
\end{equation}

\medskip
\noindent
\textbf{Chain of Inequalities and Conclusion.}

Combining all these inequalities, \eqref{eq:Uupdate}, \eqref{eq:Yupdate1}, \eqref{eq:Vupdate}, and \eqref{eq:Yupdate2}, we obtain:
\[
\begin{aligned}
  f\bigl(\bm{U}^{(t)}, \bm{\overline{Y}}^{(t)}, \bm{V}^{(t)}\bigr)
  &\;\le\;
  f\bigl(\bm{U}^{(t+1)}, \bm{\overline{Y}}^{(t+1)}, \bm{V}^{(t)}\bigr)
  \;\le\;
  f\bigl(\bm{U}^{(t+1)}, \bm{\overline{Y}}^{(t+1)}, \bm{V}^{(t+1)}\bigr)
  \\[6pt]
  &\;\le\;
  f\bigl(\bm{U}^{(t+1)}, \bm{\overline{Y}}^{(t+2)}, \bm{V}^{(t+1)}\bigr).
\end{aligned}
\]
Hence,
\[
 f\bigl(\bm{U}^{(t+1)}, \bm{\overline{Y}}^{(t+2)}, \bm{V}^{(t+1)}\bigr)
 \;\ge\;
 f\bigl(\bm{U}^{(t)}, \bm{\overline{Y}}^{(t)}, \bm{V}^{(t)}\bigr).
\]

\end{document}